\documentclass[12pt]{article}
\pdfoutput=1
\usepackage[T1]{fontenc}
\usepackage{graphicx,psfrag,epsf,color}
\usepackage{amsmath,amssymb,amsfonts}
\usepackage{physics}
\usepackage{pifont}
\usepackage{siunitx}
\usepackage{array}
\usepackage{cite}
\usepackage{rotating}
\usepackage{slashed, cancel}
\usepackage{xcolor}
\usepackage{subcaption}
\usepackage{hyperref}
\usepackage{nicefrac, physics, slashed, hyperref,titlesec,flafter,listings}
\usepackage{graphicx}
\usepackage{mathrsfs}
\usepackage{float}
\restylefloat{table}

\newcommand{\cplus}{\mathcal{C}^+}
\newcommand{\cminus}{\mathcal{C}^-}
\usepackage{tabularx}
\usepackage{booktabs}
\usepackage[margin=1.0in]{geometry}
\newcolumntype{C}{>{\centering\arraybackslash}X}
\usepackage{multirow} \usepackage{array}

\usepackage{tikz}
\usepackage[compat=1.1.0]{tikz-feynman}

\title{}
\author{andersrehult}
\date{}

\DeclareUnicodeCharacter{2212}{\ensuremath{-}}

\begin{document}

\begin{titlepage}

\vspace*{-2.0truecm}

\begin{flushright}
Nikhef-2025-006
\end{flushright}

\vspace*{1.3truecm}

\begin{center}
{
\Large \bf \boldmath Probing New Physics Through CP Violation in $B_{(s)}\to V\mu^+\mu^-$ Decays}
\end{center}
\vspace{0.9truecm}

\begin{center}
{\bf Robert Fleischer\,${}^{a,b}$,  Martijn van Hamersveld\,${}^{a}$, Tim Kortekaas\,${}^{a}$, Anders Rehult\,${}^{a,b}$, and  K. Keri Vos\,${}^{a,c}$}

\vspace{0.5truecm}

${}^a${\sl Nikhef, Science Park 105, NL-1098 XG Amsterdam,  Netherlands}

${}^b${\sl  Department of Physics and Astronomy, Vrije Universiteit Amsterdam,\\
NL-1081 HV Amsterdam, Netherlands}

{\sl $^c$Gravitational 
Waves and Fundamental Physics (GWFP),\\ 
Maastricht University, Duboisdomein 30,\\ 
NL-6229 GT Maastricht, the
Netherlands}\\[0.3cm]

\end{center}

\vspace*{1.7cm}

\begin{abstract}
\noindent
Rare decays of the kind  $B\to K^*\mu^+\mu^-$ and  $B_s\to \phi\mu^+\mu^-$ are key players for testing the Standard Model. The current experimental data for their decay rates and angular observables show tensions with the theoretical predictions that may be indications of New Physics. We present a strategy to extract the relevant short-distance coefficients in the presence of new sources of CP violation, utilizing a synergy with $B\to K\mu^+\mu^-$ decays. Using the current data as a guideline, we illustrate the new method to determine the complex coefficients $C_9^{(\prime)}$ and $C_{10}^{(\prime)}$ using only four angular observables. Interestingly, the current experimental picture leaves significant room for CP-violating New Physics. We discuss also the link to leptonic $B^0_s\to\mu^+\mu^-$ decays. We are looking forward to the implementation of these strategies at the future high-precision frontier of flavour physics. 
\end{abstract}


\vspace*{2.1truecm}

\vfill

\noindent
April 2025

\end{titlepage}


\clearpage

\thispagestyle{empty}

\vbox{}

\setcounter{page}{0}

\newpage


%
%
%
\section{Introduction}
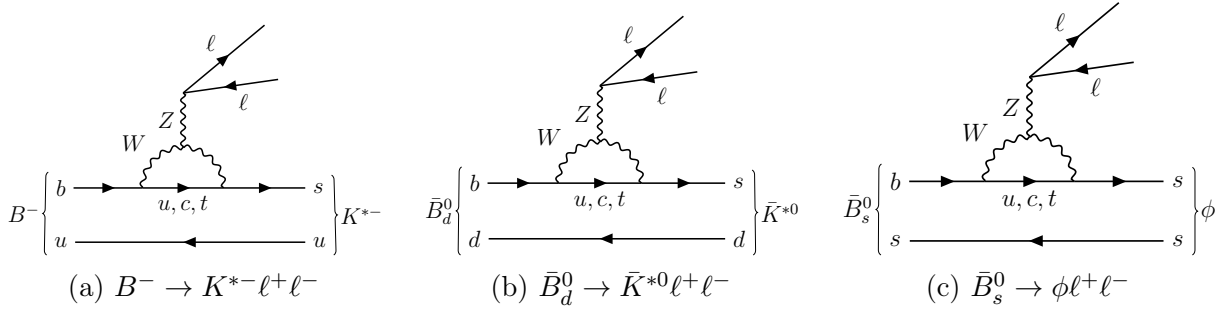
\begin{figure}[t!]
\centering
\subfloat[$B^- \to K^{\ast -} \ell^+\ell^-$]{
\resizebox{0.32\textwidth}{!}{%
\begin{tikzpicture}
\begin{feynman}

\vertex (b) {\(b\)};
\vertex[right=1.5cm of b] (Wleft);
\vertex[right=1.5cm of Wleft] (Wright);
\vertex[right=1.5cm of Wright] (s) {\(s\)};

\vertex[below=1cm of b] (uleft) {\(u\)};
\vertex[right=3cm of uleft] (ghostbottom);
\vertex[right=1.5cm of ghostbottom] (uright) {\(u\)};

\vertex[right=0.75cm of Wleft] (ghost);
\vertex[above=0.75cm of ghost] (Wanchor);

\vertex[above=1cm of Wanchor] (leptonanchor);
\vertex[right=1.5 cm of leptonanchor] (leptonghost);
\vertex[above=1.25 cm of leptonghost] (lepton1);
\vertex[right=0.25cm of leptonghost] (leptonghost2);
\vertex[above=0.25 cm of leptonghost2] (lepton2);

\vertex[left=0.20cm of Wleft] (gluonhook1);
\vertex[below=1cm of gluonhook1] (gluonghost);
\vertex[left=0.3cm of gluonghost] (gluonhook2);
\vertex[left=0.20cm of Wright] (gluonhook3);
\vertex[right=3.3cm of gluonhook2] (gluonhook4);
\vertex[above=0.4cm of gluonhook4] (gluonghost2);
\vertex[left=0.75cm of gluonghost2] (gluonhook5);
\vertex[left=1.5cm of gluonhook4] (gluonhook6);

\diagram* {
(b) -- [fermion, thick] (Wleft),
(Wleft) -- [fermion, thick, edge label'={\(u,c,t\)}] (Wright),
(Wright) -- [fermion, thick] (s),

(uright) -- [fermion, thick] (uleft),

(Wleft) -- [boson, thick,  in=180, out=90, edge label=\(W\)] (Wanchor),
(Wanchor) -- [boson, thick,  in=90, out=0] (Wright),

(Wanchor) -- [boson, thick, edge label=\(Z\)] (leptonanchor),

(leptonanchor) -- [fermion, thick, edge label={\(\ell\)}] (lepton1),
(lepton2) -- [fermion, thick, edge label={\(\ell\)}] (leptonanchor),

};
\draw [decoration={brace}, decorate] (uleft.south west) -- (b.north west)
node [pos=0.5, left] {\(B^-\)};
\draw [decoration={brace}, decorate] (s.north east) -- (uright.south east)
node [pos=0.5, right] {\(K^{\ast -}\)};
\end{feynman}
\end{tikzpicture}
}}
\hfill
\subfloat[$\bar B^0_d \to \bar K^{\ast 0} \ell^+\ell^-$]{
\resizebox{0.32\textwidth}{!}{%
\begin{tikzpicture}
\begin{feynman}

\vertex (b) {\(b\)};
\vertex[right=1.5cm of b] (Wleft);
\vertex[right=1.5cm of Wleft] (Wright);
\vertex[right=1.5cm of Wright] (s) {\(s\)};

\vertex[below=1cm of b] (uleft) {\(d\)};
\vertex[right=3cm of uleft] (ghostbottom);
\vertex[right=1.5cm of ghostbottom] (uright) {\(d\)};

\vertex[right=0.75cm of Wleft] (ghost);
\vertex[above=0.75cm of ghost] (Wanchor);

\vertex[above=1cm of Wanchor] (leptonanchor);
\vertex[right=1.5 cm of leptonanchor] (leptonghost);
\vertex[above=1.25 cm of leptonghost] (lepton1);
\vertex[right=0.25cm of leptonghost] (leptonghost2);
\vertex[above=0.25 cm of leptonghost2] (lepton2);

\vertex[left=0.20cm of Wleft] (gluonhook1);
\vertex[below=1cm of gluonhook1] (gluonghost);
\vertex[left=0.3cm of gluonghost] (gluonhook2);
\vertex[left=0.20cm of Wright] (gluonhook3);
\vertex[right=3.3cm of gluonhook2] (gluonhook4);
\vertex[above=0.4cm of gluonhook4] (gluonghost2);
\vertex[left=0.75cm of gluonghost2] (gluonhook5);
\vertex[left=1.5cm of gluonhook4] (gluonhook6);

\diagram* {
(b) -- [fermion, thick] (Wleft),
(Wleft) -- [fermion, thick, edge label'={\(u,c,t\)}] (Wright),
(Wright) -- [fermion, thick] (s),

(uright) -- [fermion, thick] (uleft),

(Wleft) -- [boson, thick,  in=180, out=90, edge label=\(W\)] (Wanchor),
(Wanchor) -- [boson, thick,  in=90, out=0] (Wright),

(Wanchor) -- [boson, thick, edge label=\(Z\)] (leptonanchor),

(leptonanchor) -- [fermion, thick, edge label={\(\ell\)}] (lepton1),
(lepton2) -- [fermion, thick, edge label={\(\ell\)}] (leptonanchor),

};
\draw [decoration={brace}, decorate] (uleft.south west) -- (b.north west)
node [pos=0.5, left] {\(\bar B^0_d\)};
\draw [decoration={brace}, decorate] (s.north east) -- (uright.south east)
node [pos=0.5, right] {\(\bar K^{\ast 0}\)};
\end{feynman}
\end{tikzpicture}
}}
\hfill
\subfloat[$\bar B^0_s \to \phi \ell^+\ell^-$]{
\resizebox{0.32\textwidth}{!}{%
\begin{tikzpicture}
\begin{feynman}

\vertex (b) {\(b\)};
\vertex[right=1.5cm of b] (Wleft);
\vertex[right=1.5cm of Wleft] (Wright);
\vertex[right=1.5cm of Wright] (s) {\(s\)};

\vertex[below=1cm of b] (uleft) {\(s\)};
\vertex[right=3cm of uleft] (ghostbottom);
\vertex[right=1.5cm of ghostbottom] (uright) {\(s\)};

\vertex[right=0.75cm of Wleft] (ghost);
\vertex[above=0.75cm of ghost] (Wanchor);

\vertex[above=1cm of Wanchor] (leptonanchor);
\vertex[right=1.5 cm of leptonanchor] (leptonghost);
\vertex[above=1.25 cm of leptonghost] (lepton1);
\vertex[right=0.25cm of leptonghost] (leptonghost2);
\vertex[above=0.25 cm of leptonghost2] (lepton2);

\vertex[left=0.20cm of Wleft] (gluonhook1);
\vertex[below=1cm of gluonhook1] (gluonghost);
\vertex[left=0.3cm of gluonghost] (gluonhook2);
\vertex[left=0.20cm of Wright] (gluonhook3);
\vertex[right=3.3cm of gluonhook2] (gluonhook4);
\vertex[above=0.4cm of gluonhook4] (gluonghost2);
\vertex[left=0.75cm of gluonghost2] (gluonhook5);
\vertex[left=1.5cm of gluonhook4] (gluonhook6);

\diagram* {
(b) -- [fermion, thick] (Wleft),
(Wleft) -- [fermion, thick, edge label'={\(u,c,t\)}] (Wright),
(Wright) -- [fermion, thick] (s),

(uright) -- [fermion, thick] (uleft),

(Wleft) -- [boson, thick,  in=180, out=90, edge label=\(W\)] (Wanchor),
(Wanchor) -- [boson, thick,  in=90, out=0] (Wright),

(Wanchor) -- [boson, thick, edge label=\(Z\)] (leptonanchor),

(leptonanchor) -- [fermion, thick, edge label={\(\ell\)}] (lepton1),
(lepton2) -- [fermion, thick, edge label={\(\ell\)}] (leptonanchor),

};
\draw [decoration={brace}, decorate] (uleft.south west) -- (b.north west)
node [pos=0.5, left] {\(\bar B^0_s\)};
\draw [decoration={brace}, decorate] (s.north east) -- (uright.south east)
node [pos=0.5, right] {\(\phi\)};
\end{feynman}
\end{tikzpicture}
}}
\caption{Leading-order SM Feynman diagrams for $B_{(s)} \to K^\ast(\phi)\ell^+\ell^-$ decays.}
\label{fig:diagrams_btovll}
\end{figure}

Decays of $B$ mesons originating from $b \to s\ell^+\ell^-$ quark-level transitions, with $\ell= e, \mu, \tau$, are rare flavour-changing neutral current processes. These decays play an important role in testing the Standard Model of particle physics (SM). We are living in exciting times as experimental data for observables provided by such modes show tensions with SM predictions. A particularly interesting probe is offered by CP-violating phenomena. In a previous paper \cite{Fleischer:2022klb}, we have developed a strategy to ``fingerprint" CP-violating New Physics (NP) effects with decays of the kind $B\to K\mu^+\mu^-$. In the present study, we will focus on the exclusive channels
$B^- \to K^{\ast -} \mu^+\mu^-$, $\bar B^0_d \to \bar K^{0 \ast} \mu^+\mu^-$ and $\bar B^0_s \to \phi \mu^+\mu^-$ with their CP-conjugates.
In Fig.~\ref{fig:diagrams_btovll}, we show the Feynman diagrams arising in the SM for these decay processes which differ only by their spectator quarks. We will refer to these channels generically as $B_{(s)} \to V\ell^+\ell^-$ modes. 

In comparison with the charged and neutral $B \to K\ell^+\ell^-$ decays studied in Ref.~\cite{Fleischer:2022klb}, channels with a vector meson in the final state provide access to a wider range of observables. The non-zero spin of the final-state meson introduces an additional angular degree of freedom, resulting in 12 terms in the angular distribution instead of the three found in $B \to K\ell^+\ell^-$ decays. Although not all of these observables are independent from one another, as we will discuss below in more detail, the enlarged set offers various additional probes for exploring NP contributions to $b \to s\ell^+\ell^-$ processes. As in the case of the $B \to K\ell^+\ell^-$ modes,
the differential decay rates and observables depend on the four-momentum transfer $q^2$ to the $\ell^+\ell^-$ pair. Observables in $B_{(s)} \to K^{\ast}(\phi)\mu^+\mu^-$ decays have extensively been studied in experimental analyses (see, for instance,  \cite{LHCb:2021xxq,LHCb:2023gpo,CMS-PAS-BPH-21-002,CMS:2024atz,LHCb:2021zwz,LHCb:2014cxe,ATLAS:2018gqc,LHCb:2015svh,CMS:2015bcy,Belle:2016fev,CMS:2017rzx,LHCb:2020lmf}). These analyses have revealed several discrepancies with respect to the SM predictions (sometimes referred to as ``anomalies" in the literature): the measured CP-averaged $B_s \to \phi\mu^+\mu^-$ branching ratio lies about $4 \sigma$ below the SM prediction \cite{Fleischer:2024fkm, Gubernari:2022hxn,Bharucha:2015bzk,Straub:2018kue}, and the optimized angular observable $P_5^\prime$ \cite{Descotes-Genon:2012isb}, designed to be robust with respect to hadronic SM effects, is found to be $(2$-$4) \sigma$ lower than the SM predictions, depending on the considered $q^2$ bins \cite{LHCb:2020lmf,CMS-PAS-BPH-21-002,CMS:2024atz,Gubernari:2022hxn,Straub:2018kue}.

The semileptonic $B_{(s)} \to V\ell^+\ell^-$ decays are described theoretically through a low-energy effective Hamiltonians, allowing -- in addition to the SM contributions --  a model-independent description of NP effects which are encoded in the short-distance (Wilson) coefficients of local four-fermion operators. The key coefficients for the decays at hand are $C_9^{(\prime)}$ and $C_{10}^{(\prime)}$. In the presence of CP-violating effects, these coefficients are complex quantities. It is useful to introduce the following combinations:
\begin{align}\label{eq:pmdef_vector}
    \mathcal{C}_i^\pm &\equiv C_i \pm C_i^\prime \ , \quad i \in \{9,10\} \ .
\end{align}
Both the $B_{(s)} \to V\ell^+\ell^-$ and $B \to K\ell^+\ell^-$ decays are described by the same Hamiltonian. Different decay amplitudes result from the different hadronic matrix elements between the initial and final states. Due to the pseudo-scalar and vector mesons in the final states, a key difference between the $B \to K\ell^+\ell^-$ and $B_d \to K^\ast \ell^+\ell^-$ decays lies in their dependencies on the Wilson coefficients $C_9^{(\prime)}$ and $C_{10}^{(\prime)}$. The $B \to K\ell^+\ell^-$ decay depends exclusively on the combinations $\cplus_9$ and $\cplus_{10}$, whereas the $B_d \to K^\ast\ell^+\ell^-$ decay depends on these combinations as well as on $\cminus_9$ and $\cminus_{10}$ \cite{Altmannshofer:2008dz}.

In the following discussion, we will focus on $\bar B^0_d \to \bar K^{* 0}\mu+\mu-$ and its CP conjugate as our main ``working example", collectively denoted as $B_d \to K^{\ast} \mu^+\mu^-$. Previously, measurements of CP-averaged observables in the $B_d \to K^\ast \mu^+\mu^-$ decay have been used to determine the Wilson coefficients $\cplus_9$, $\cplus_{10}$, $\cminus_9$ and $\cminus_{10}$, under the assumption that these coefficients are real \cite{LHCb:2024onj}. Our goal is to relax this assumption and determine the coefficients in a transparent way while allowing them to be complex. To this end, we will construct a minimal basis of observables, including both CP-averaged and CP-violating observables. Moreover,
we will utilize the method for the $B\to K\mu^+\mu^-$ case presented in Ref.~\cite{Fleischer:2022klb}, allowing the determination of the
coefficients $\cplus_9$ and $\cplus_{10}$ from CP-violating effects. Using these combinations of Wilson coefficients as an input, we can extract $\cminus_9$ and $\cminus_{10}$ from the $B_d \to K^\ast \mu^+\mu^-$ observables. In this way, we finally obtain a transparent picture of possible CP-violating NP contributions to $C_9$, $C_9^\prime$, $C_{10}$, and $C_{10}^\prime$. The Wilson coefficient combination $\cminus_{10}$ has an immediate and important application for analyses of (pseudo)-scalar NP effects in the leptonic $B^0_s\to\mu^+\mu^-$ decay, as discussed in detail in Refs.~\cite{Fleischer:2017yox,Fleischer:2024fkm}.

The outline of this paper is as follows: In Section~\ref{sec:theo}, we introduce the relevant theoretical framework with the effective Hamiltonian, angular distribution, hadronic and long-distance parameters and observables. In Section~\ref{sec:data-theory}, we discuss the current picture resulting from the experimental data and make a comparison with the theoretical predictions. We move then on to discuss the sensitivity of the $B_{(s)} \to V\ell^+\ell^-$ observables to NP effects in Section~\ref{sec:sens-obs}, where we present our main strategy. In Section~\ref{sec:sec-extr}, we apply it to extract the complex coefficients $\cminus_9$ and $\cminus_{10}$, illustrating also new perspectives and opportunities through future scenarios. Finally, we summarize our conclusions in Section~\ref{sec:concl}.

\section{Theoretical Framework}
\label{sec:theo}
\subsection{Effective Hamiltonian}
The general low-energy effective Hamiltonian for $b \to s \ell^+\ell^-$ transitions reads \cite{Descotes-Genon:2020tnz, Altmannshofer:2008dz, Gratrex:2015hna,Buchalla:1995vs} 
\begin{equation}\label{eq:ham}
    \mathcal{H}_{\rm eff} = - \frac{4 G_F}{\sqrt{2}} \left[\lambda_u \Big\{C_1 (\mathcal{O}_1^c - \mathcal{O}_1^u) + C_2 (\mathcal{O}_2^c - \mathcal{O}_2^u)\Big\} + \lambda_t \sum\limits_{i \in I} C_i \mathcal{O}_i \right] \ ,
\end{equation}
where $\lambda_q = V_{qb} V_{qs}^*$ and $I = \{1c, 2c, 3, 4, 5, 6, 8, 7^{(\prime)}, 9^{(\prime)}\ell, 10^{(\prime)}\ell, S^{(\prime)}\ell, P^{(\prime)}\ell, T^{(\prime)}\ell\}$. The terms proportional to $\lambda_u$ are doubly-Cabibbo suppressed, i.e., contribute at the $\mathcal{O}(\lambda^2) \sim 5\%$ level, where $\lambda\equiv |V_{us}|\sim0.22$ \cite{Wolfenstein:1983yz,Buras:1994ec}. In the following, we neglect these suppressed terms. The operators are given as
\begin{equation}
    \begin{aligned}
    \mathcal{O}_{7^{(\prime)}} &= \frac{e}{(4\pi)^2} m_b [\bar s \sigma^{\mu\nu} P_{R(L)} b] F_{\mu\nu}, & \mathcal{O}_{S^{(\prime)}} &= \frac{e^2}{(4\pi)^2} m_b [\bar s P_{R(L)} b] (\bar \ell \ell),\\
    \mathcal{O}_{9^{(\prime)}} &= \frac{e^2}{(4\pi)^2} [\bar s \gamma^\mu P_{L(R)} b] (\bar \ell \gamma_\mu \ell), & \mathcal{O}_{P^{(\prime)}} &= \frac{e^2}{(4\pi)^2} m_b [\bar s P_{R(L)} b] (\bar \ell \gamma_5 \ell),\\
    \mathcal{O}_{10^{(\prime)}} &= \frac{e^2}{(4\pi)^2} [\bar s \gamma^\mu P_{L(R)} b] (\bar \ell \gamma_\mu \gamma_5 \ell), & \mathcal{O}_{T^{(\prime)}} &= \frac{e^2}{(4\pi)^2} [\bar s \sigma^{\mu\nu} P_{R(L)} b] (\bar \ell \sigma_{\mu\nu} ) \ ,
    \end{aligned}
\end{equation}
with $P_{R(L)} = \frac{1}{2} (1 \pm \gamma_5)$ and $\sigma_{\mu\nu} = \frac{i}{2}[\gamma_\mu, \gamma_\nu]$. For simplicity, we have omitted an index $\ell$ on the operators. For our numerical analysis, we focus on decays with $\ell = \mu$, however the formalism below holds also for $\ell=e,\tau$. 

In the following, we neglect (pseudo)-scalar NP contributions. As discussed in detail in \cite{Fleischer:2024fkm}, such NP contributions lift the helicity suppression in $B_s \to \mu^+\mu^-$. Consequently, measurements of the $B_s \to \mu^+\mu^-$ branching ratio place strong constraints on the impact of muonic (pseudo)-scalar NP contributions, rendering their effects on semileptonic $b \to s \mu^+\mu^-$ decays negligibly small. We discuss $B_s\to \mu^+\mu^-$ briefly in Sec.~\ref{sec:bsmumu}. In addition, we also neglect tensor coefficients.

For the SM Wilson coefficients, we use \cite{EOSAuthors:2021xpv} 
\begin{equation}
    C_7^{\rm SM} = -0.33, \quad C_9^{\rm SM} = 4.27, \quad C_{10}^{\rm SM} = -4.17 \ ,
\end{equation}
which are flavour universal (i.e.\ equal for $\ell=\mu=e=\tau$) and determined at $\mu=m_b$. In addition, we use the quark masses in the ${\overline{\rm MS}}$ scheme at $m_b$: $\bar{m}_b(m_b) = 4.18\pm 0.03$ and $\bar{m}_s(m_b)= 0.078\pm 0.007$ \cite{ParticleDataGroup:2020ssz}. As mentioned in the introduction, it is convenient to define sums and differences of the Wilson coefficients and their primed counterparts according to \eqref{eq:pmdef_vector}.

\subsection{Angular distribution}
The angular distribution for $B_{(s)}\to V\ell^+\ell^-$ decays then reads \cite{Descotes-Genon:2015hea,Bobeth:2017vxj,Altmannshofer:2008dz}:
\begin{align} \label{eq:angular_distr_btovll}
   \frac{d^4\Gamma(B_{(s)} \to V \ell^+\ell^-)}{dq^2 d\cos\theta_M d\cos\theta_\ell d\phi} &= \frac{9 }{32\pi}\left[ J_{1s} \sin^2\theta_M +J_{1c}\cos^2\theta_M + J_{2s} \sin^2\theta_M\cos2\theta_\ell \right. \nonumber\\
   & \left. + J_{2c} \cos^2\theta_M \cos2\theta_\ell + J_3 \sin^2\theta_M \sin^2\theta_\ell \cos2\phi+ J_4 \sin2\theta_M \sin2\theta_\ell \cos\phi \right. \nonumber \\
   & \left. + J_5 \sin2\theta_M \sin\theta_\ell \cos\phi + J_{6s}\sin^2\theta_M\cos\theta_\ell + J_{6c}\cos^2\theta_M\cos\theta_\ell \right. \nonumber \\
   & \left. + J_7 \sin2\theta_M \sin\theta_\ell \sin\phi + J_8 \sin2\theta_M \sin\phi + J_9 \sin^2\theta_M \sin^2\theta_\ell \sin2\phi  \right] \ ,
\end{align}
where $\theta_M, \theta_\ell$ and $\phi$ are kinematical angles defined in terms of the four-momenta of the decay products arising in the $B(p_B) \to V[\to M_1(p_1) M_2(p_2] \ell^+(p_+) \ell^-(p_1)$ decay chains \cite{Descotes-Genon:2015hea,Kruger:1999xa}: $\theta_M$ is the polar angle between $\Vec p_B$ and $\Vec p_1$ in the rest frame of the $V$ meson, $\theta_\ell$ is the polar angle between $\Vec p_B$ and $\Vec p_+$ in the dilepton rest frame, and $\phi$ is the angle between the planes generated by $(\Vec p_1, \Vec p_2)$ and $(\Vec p_+, \Vec p_-)$ in the $B$-meson rest frame. For simplicity, we do not explicitly give the $q^2$ dependence of the angular coefficients $J_i \equiv J_i(q^2)$.

The angular coefficients can be expressed in the following form \cite{Altmannshofer:2008dz,Descotes-Genon:2015hea}:
\begin{equation}\label{eq:j_coefficients}
\begin{aligned}
 J_{1 s}&=\frac{\left(2+\beta_{\ell}^{2}\right)}{4}\left[\left|\mathcal{A}_{\perp}^{L}\right|^{2}+\left|\mathcal{A}_{\|}^{L}\right|^{2}+\left|\mathcal{A}_{\perp}^{R}\right|^{2}+\left|\mathcal{A}_{\|}^{R}\right|^{2}\right]+\frac{4 m_{\ell}^{2}}{q^2} \operatorname{Re}\left(\mathcal{A}_{\perp}^{L} \mathcal{A}_{\perp}^{R^{*}}+\mathcal{A}_{\|}^{L} \mathcal{A}_{\|}^{R^{*}}\right)\ , \\
 J_{1 c}&=\left|\mathcal{A}_{0}^{L}\right|^{2}+\left|\mathcal{A}_{0}^{R}\right|^{2}+\frac{4 m_{\ell}^{2}}{q^2}\left[\left|\mathcal{A}_{t}\right|^{2}+2 \operatorname{Re}\left(\mathcal{A}_{0}^{L} \mathcal{A}_{0}^{R^{*}}\right)\right]+\beta_{\ell}^{2}\left|\mathcal{A}_{S}\right|^{2}\ , \\
 J_{2 s}&=\frac{\beta_{\ell}^{2}}{4}\left[\left|\mathcal{A}_{\perp}^{L}\right|^{2}+\left|\mathcal{A}_{\|}^{L}\right|^{2}+\left|\mathcal{A}_{\perp}^{R}\right|^{2}+\left|\mathcal{A}_{\|}^{R}\right|^{2}\right]\qc\qquad  J_{2 c}=-\beta_{\ell}^{2}\left[\left|\mathcal{A}_{0}^{L}\right|^{2}+\left|\mathcal{A}_{0}^{R}\right|^{2}\right]\ , \\
 J_{3}&=\frac{1}{2} \beta_{\ell}^{2}\left[\left|\mathcal{A}_{\perp}^{L}\right|^{2}-\left|\mathcal{A}_{\|}^{L}\right|^{2}+\left|\mathcal{A}_{\perp}^{R}\right|^{2}-\left|\mathcal{A}_{\|}^{R}\right|^{2}\right]\qc\qquad J_{4}=\frac{1}{\sqrt{2}} \beta_{\ell}^{2}\left[\operatorname{Re}\left(\mathcal{A}_{0}^{L} \mathcal{A}_{\|}^{L^{*}}+\mathcal{A}_{0}^{R} \mathcal{A}_{\|}^{R^{*}}\right)\right]\ , \\
 J_{5}&=\sqrt{2} \beta_{\ell}\left[\operatorname{Re}\left(\mathcal{A}_{0}^{L} \mathcal{A}_{\perp}^{L^{*}}-\mathcal{A}_{0}^{R} \mathcal{A}_{\perp}^{R^{*}}\right)-\frac{m_{\ell}}{\sqrt{q^2}} \operatorname{Re}\left(\mathcal{A}_{\|}^{L} \mathcal{A}_{S}^{*}+\mathcal{A}_{\|}^{R^{*}} \mathcal{A}_{S}\right)\right]\ , \\
 J_{6 s}&=2 \beta_{\ell}\left[\operatorname{Re}\left(\mathcal{A}_{\|}^{L} \mathcal{A}_{\perp}^{L^{*}}-\mathcal{A}_{\|}^{R} \mathcal{A}_{\perp}^{R^{*}}\right)\right]\qc\qquad\qquad\quad~~J_{6 c}=4 \beta_{\ell} \frac{m_{\ell}}{\sqrt{q^2}} \operatorname{Re}\left(\mathcal{A}_{0}^{L} \mathcal{A}_{S}^{*}+\mathcal{A}_{0}^{R^{*}} \mathcal{A}_{S}\right)\ , \\
 J_{7}&=\sqrt{2} \beta_{\ell}\left[\operatorname{Im}\left(\mathcal{A}_{0}^{L} \mathcal{A}_{\|}^{L^{*}}-\mathcal{A}_{0}^{R} \mathcal{A}_{\|}^{R^{*}}\right)+\frac{m_{\ell}}{\sqrt{q^2}} \operatorname{Im}\left(\mathcal{A}_{\perp}^{L} \mathcal{A}_{S}^{*}-\mathcal{A}_{\perp}^{R^{*}} \mathcal{A}_{S}\right)\right]\ , \\
 J_{8}&=\frac{1}{\sqrt{2}} \beta_{\ell}^{2}\left[\operatorname{Im}\left(\mathcal{A}_{0}^{L} \mathcal{A}_{\perp}^{L^{*}}+\mathcal{A}_{0}^{R} \mathcal{A}_{\perp}^{R^{*}}\right)\right]\qc\qquad\qquad\quad J_9=\beta_{\ell}^{2}\left[\operatorname{Im}\left(\mathcal{A}_{\|}^{L^{*}} \mathcal{A}_{\perp}^{L}+\mathcal{A}_{\|}^{R^{*}} \mathcal{A}_{\perp}^{R}\right)\right]\ ,
\end{aligned}
\end{equation}
where $\beta_{\ell} \equiv \sqrt{1- 4m_{\ell}^{2}/q^2}$ and $m_\ell$ is the lepton mass. 

The amplitudes $\mathcal{A}_{\lambda}^{L,R}$ where $\lambda=0,\perp,\parallel, t$ are transversity amplitudes which can be expressed as follows \cite{Bobeth:2012vn, Bobeth:2017vxj,LHCb:2023gel}:
\begin{equation}\label{eq:transversity_amplitudes}
\begin{aligned}
\mathcal{A}_{\perp}^{L, R}(q^2) &= \mathcal{N}(q^2)\Big\{ \Big[ \left( \mathcal{C}_9 + \mathcal{C}_9^{\prime} \right) 
\mp \left(\mathcal{C}_{10} + \mathcal{C}_{10}^{\prime}\right)\Big] 
\mathcal{F}_{\perp}\left(q^2\right)  \\ 
& \quad + \frac{2 m_b M_B}{q^2} \Big[\left(\mathcal{C}_7 + \mathcal{C}_7^{\prime}\right) 
\mathcal{F}_{\perp}^T\left(q^2\right) - 16 \pi^2 \frac{M_B}{m_b} 
\mathcal{H}_{\perp}\left(q^2\right)\Big] \Big\} \ , \\
\mathcal{A}_{\|}^{L, R}(q^2) &= -\mathcal{N}(q^2)\Big\{ \Big[ \left( \mathcal{C}_9 - \mathcal{C}_9^{\prime} \right) 
\mp \left(\mathcal{C}_{10} - \mathcal{C}_{10}^{\prime}\right)\Big] 
\mathcal{F}_{\|}\left(q^2\right)  \\ 
& \quad + \frac{2 m_b M_B}{q^2} \Big[\left(\mathcal{C}_7 - \mathcal{C}_7^{\prime}\right) 
\mathcal{F}_{\|}^T\left(q^2\right) - 16 \pi^2 \frac{M_B}{m_b} 
\mathcal{H}_{\|}\left(q^2\right)\Big] \Big\} \ , \\
\mathcal{A}_{0}^{L, R}(q^2) &= -\mathcal{N}(q^2) \frac{M_B}{\sqrt{q^2}}\Big\{ \Big[ \left( \mathcal{C}_9 - \mathcal{C}_9^{\prime} \right) 
\mp \left(\mathcal{C}_{10} - \mathcal{C}_{10}^{\prime}\right)\Big] 
\mathcal{F}_{0}\left(q^2\right)  \\ 
& \quad + \frac{2 m_b M_B}{q^2} \Big[\left(\mathcal{C}_7 - \mathcal{C}_7^{\prime}\right) 
\mathcal{F}_{0}^T\left(q^2\right) - 16 \pi^2 \frac{M_B}{m_b} 
\mathcal{H}_{0}\left(q^2\right)\Big] \Big\} \ , \\
\mathcal{A}_t(q^2) &= -2 \mathcal{N}(q^2)\left(\mathcal{C}_{10}-\mathcal{C}_{10}^{\prime}\right) \mathcal{F}_t\left(q^2\right) \ , \quad \mathcal{A}_S(q^2) = 0 \ ,
\end{aligned}
\end{equation}
where
\begin{equation}
\mathcal{N}(q^2) = G_F \alpha_e V_{t b} V_{t s}^* \sqrt{\frac{q^2 \beta_\ell \sqrt{\lambda_B}}{3 \cdot 2^{10} \pi^5 M_B}} \ ,
\end{equation}
and $\lambda_B \equiv \lambda(q^2, M_B^2, M_V^2) = q^4 + M_B^4 + M_V^4 - 2q^2 M_B^2 - 2q^2 M_V^2 - 2 M_B^2 M_V^2$  is the Källén function. Here, $\mathcal{F}^{(T)}_{\perp, \parallel,0,t}(q^2)$ and $\mathcal{H}_{\perp, \parallel, 0}$ parametrize the ``local'' and ``non-local'' long-distance effects, respectively. We discuss the latter in the next subsection. 

The $\mathcal{F}^{(T)}_{\perp, \parallel,0,t}(q^2)$ functions are hadronic form factors parametrized as \cite{LHCb:2023gel}
\begin{equation}\label{eq:ff_conversion}
\begin{aligned}
    \mathcal{F}_\perp &= \frac{\sqrt{2 \lambda_B}}{M_B (M_B + M_V)} V \ , 
   & \mathcal{F}_\parallel &= \frac{\sqrt{2} (M_B + M_V)}{M_B} A_1 \ , \\
    \mathcal{F}_0 &= \frac{(M_B^2 - q^2 - M_V^2) (M_B + M_V)^2 A_1 - \lambda_B A_2}{2 M_V M_B^2 (M_B + M_V)} \ ,&  \mathcal{F}_\perp^T &= \frac{\sqrt{2 \lambda_B}}{M_B^2} T_1 \ ,\\
    \mathcal{F}_\parallel^T &= \frac{\sqrt{2} (M_B^2 - M_V^2)}{M_B^2} T_2 \ ,&  \mathcal{F}_t &= \frac{\sqrt{\lambda_B}}{M_B \sqrt{q^2}} A_0 \ , \\
    \mathcal{F}_0^T &= \frac{q^2 (M_B^2 + 3 M_V^2 - q^2)}{2 M_B^3 M_V} T_2 - \frac{q^2 \lambda_B}{2 M_B^3 M_V (M_B^2 - M_V^2)} T_3 \ ,
\end{aligned}
\end{equation}
where  \cite{Bobeth:2010wg,Ball:2004rg}:
\begin{equation}\label{eq:btov_local_ffs}
\begin{aligned}
    \mel{V(k,\epsilon)}{\bar q \gamma_\mu b}{B(p)} &= \frac{2V(q^2)}{M_B + M_V} \varepsilon_{\mu \rho \sigma \tau} \epsilon^{\ast \rho} p^\sigma k^\tau \ ,\\
    \mel{V(k,\epsilon)}{\bar q \gamma_\mu \gamma_5 b}{B(p)} &= i \epsilon^{\ast \rho} \bigg[ 2 M_V A_0(q^2) \frac{q_\mu q_\rho}{q^2} + (M_B + M_V) A_1(q^2) \bigg( g_{\mu \rho} - \frac{q_\mu q_\rho}{q^2} \bigg)\\
    &- A_2(q^2) \frac{q_\rho}{M_B + M_V} \bigg( (p + k)_\mu - \frac{M_B^2 - M_V^2}{q^2} (p - k)_\mu \bigg)\bigg] \ ,\\
     \mel{V(k,\epsilon)}{\bar q \sigma_{\mu \nu} q^\nu b}{B(p)} &= -2 T_1(q^2) \varepsilon_{\mu \rho \sigma \tau} \epsilon^{\ast \rho} p^\sigma k^\tau \ ,\\
    \mel{V(k,\epsilon)}{\bar q \sigma_{\mu \nu} \gamma_5 q^\nu b}{B(p)} &= i T_2(q^2) (\epsilon_\mu^\ast (M_B^2 - M_V^2) - (\epsilon^\ast \cdot q) (p + k)_\mu)\\
    &+ i T_3(q^2) (\epsilon^\ast \cdot q) \bigg( q_\mu - \frac{q^2}{M_B^2 - M_V^2} (p + k)_\mu \bigg) \ .
\end{aligned}
\end{equation}
Here $\varepsilon_{\mu \rho \sigma \tau}$ is the Levi-Civita tensor, $p$ and $k$ are the four-momenta of the $B$ and $V$ mesons, respectively, $q \equiv p - k$, and $\epsilon$ is the polarization vector of the $V$ meson. 

These form factors can be obtained using light-cone sum rule techniques and using lattice QCD. For our numerical analysis we use the results of \cite{Bharucha:2015bzk}. 

\subsection{Non-Local, Long-Distance Parameters}
The $\mathcal{H}_{\perp,\parallel,0}$ in \eqref{eq:transversity_amplitudes}, which parametrize the non-local, long-distance hadronic effects, arise from higher-order contributions from the current-current operators $O_1^c$ and $O_2^c$ in the effective Hamiltonian. In the case of $B_{(s)} \to V \mu^+\mu^-$ decays, these effects have been addressed using various theoretical and experimental approaches. A perturbative method was discussed in~\cite{Grinstein:2004vb}, applicable in regions of the $q^2$ spectrum far from the $c \bar c$ resonances. Approaches that model these effects through light-cone sum rules are found in~\cite{Khodjamirian:2010vf,Khodjamirian:2012rm}. Perturbative QCD methods exploiting the analytic structure of the relevant correlation functions are discussed in~\cite{Bobeth:2017vxj,Gubernari:2020eft,Gubernari:2022hxn}. Experimentally, the LHCb collaboration has performed analyses aimed at disentangling long-distance contributions from short-distance NP effects described by Wilson coefficients (assuming no new sources of CP violation) \cite{LHCb:2024onj,LHCb:2023gpo}. Despite interesting progress in this area, the treatment of the long-distance effects and their associated uncertainties remain under discussion. Here especially data-driven methods seem promising to further our understanding of these QCD effects.

The goal of this paper is to outline a strategy for extracting and disentangling possible complex NP Wilson coefficients. As such we refrain from discussing the long-distance effects in detail and use the long-distance model of \cite{Bobeth:2017vxj}. Compared to the perturbative model, this description includes more detailed effects from $c \bar c$ resonances, which will be important in our study of CP asymmetries. At the same time, as we also detail below, this gives a degree of model-dependence to our predictions for the CP asymmetries. 

\subsection{Observables}
\subsubsection{CP-Averaged Observables}
Integrating the differential angular distribution in \eqref{eq:angular_distr_btovll} over $\cos\theta_M \in [-1,1]$, $\cos\theta_\ell \in [-1,1]$ and $\phi \in [0,2\pi]$ yields the following differential decay rate:
\begin{equation}
    \frac{d\Gamma(B_{(s)} \to V \ell^+\ell^-)}{dq^2} = \frac{8}{32\pi}\left[ J_{1s} + J_{1c} + J_{2c}+ J_{2s} \right] \ \ .
\end{equation}

The expressions for the CP-conjugate modes $\bar B_{(s)} \to \bar V \ell^+\ell^-$ can be obtained by replacing 
\begin{equation}\label{eq:j_signs}
    J_{1,2,3,4,7} \to \bar J_{1,2,3,4,7} \ , \quad J_{5,6,8,9} \to - \bar J_{5,6,8,9} \ ,
\end{equation}
where $\bar{J}$ is the CP-conjugate of $J$. Specifically, $\bar{J}$ is obtained by taking the complex-conjugate of the CP-violating phases, both the CKM phases entering through $\mathcal{N}$ and possible the NP phases of the Wilson coefficients. The minus signs are a consequence of the angular convention (see, e.g.,\cite{Altmannshofer:2008dz,Descotes-Genon:2015hea}). 

The CP-averaged differential branching ratio 
\begin{equation}
\begin{aligned}
    \frac{d\mathcal{B}}{dq^2} &\equiv \frac{\tau_B}{2} \bigg[\frac{d\Gamma(B_{(s)} \to V \ell^+\ell^-)}{dq^2} + \frac{d\Gamma(\bar B_{(s)} \to \bar V \ell^+\ell^-)}{dq^2} \bigg] \  \end{aligned}
\end{equation}
can then be written as 
\begin{equation}
\begin{aligned}
\frac{d\mathcal{B}}{dq^2}     &= \frac{\tau_B}{2} \frac{8}{32\pi}\left[ J_{1s} + \bar J_{1s} + J_{1c} + \bar J_{1c} + J_{2c} + \bar J_{2c} + J_{2s} + \bar J_{2s} \right] \ \ .
\end{aligned}
\end{equation}
For our studies, it is also interesting to consider the branching ratio integrated over $q^2$. Following the LHCb convention (see, e.g., \cite{LHCb:2016ykl,LHCb:2020lmf}), we normalize this quantity by the $q^2$ bin width, yielding
\begin{equation}
    \mathcal{B}[q^2_{\rm min},q^2_{\rm max}] \equiv \frac{1}{q^2_{\rm min}-q^2_{\rm max}} \int\limits_{q^2_{\rm min}}^{q^2_{\rm max}} \frac{d \mathcal{B}}{dq^2} dq^2 \ .
\end{equation}

In addition to the CP-averaged branching ratio, each angular coefficient can be measured separately, though they are not all independent from each other.

Following \cite{Descotes-Genon:2013vna}, we consider the longitudinal polarization fraction $F_L$ and the optimized angular observables $P_i^{(\prime)}$ first discussed in \cite{Descotes-Genon:2012isb}. However, to facilitate easy comparison with the experimental data, we use the LHCb definition given in \cite{LHCb:2015svh,LHCb:2013ghj,LHCb:2020lmf}:
\begin{align}\label{eq:btovll_cpavg_obs}
    F_L &\equiv S_{1c} \ , \quad 
    P_1 \equiv \frac{2 S_3}{(1 - F_L)} \ , \quad 
    P_2 \equiv \frac{1}{2} \frac{S_{6s}}{(1 - F_L)} \ , \quad
    P_3 \equiv \frac{-S_9}{(1 - F_L)} \ , \notag \\
    P_4^\prime &\equiv \frac{S_4}{\sqrt{F_L(1 - F_L)}} \ , \quad
    P_5^\prime \equiv \frac{S_5}{\sqrt{F_L(1 - F_L)}} \ , \quad
    P_6^\prime \equiv \frac{S_7}{\sqrt{F_L(1 - F_L)}} \ .
\end{align}
Here the $S_i$ are defined as \cite{Altmannshofer:2008dz}
\begin{equation}
    S_i[q^2_{\rm min},q^2_{\rm max}] \equiv \frac{\int_{q^2_{\rm min}}^{q^2_{\rm max}} \frac{d \left(J_i + \bar J_i \right)}{dq^2} dq^2}{\int_{q^2_{\rm min}}^{q^2_{\rm max}} \frac{d \left(\Gamma +  \bar\Gamma \right)}{dq^2}dq^2} \ .
\end{equation}

\subsubsection{CP-Violating Observables}
Besides the CP-conserving observables, additional information can be obtained from the direct CP asymmetries given by
\begin{equation}
        \mathcal{A}_{\rm CP}^{\rm dir} = \frac{\Gamma(B_{(s)} \to V \ell^+\ell^-) - \Gamma(\bar B_{(s)} \to \bar V \ell^+\ell^-)}{\Gamma(B_{(s)} \to V \ell^+\ell^-) + \Gamma(\bar B_{(s)} \to \bar V \ell^+\ell^-)} \ .
\end{equation}
In addition, by taking the difference of the $J_i$, we can construct CP asymmetries for each angular mode given by \cite{Altmannshofer:2008dz}
\begin{equation}\label{eq:angular_cp_asymmetries}
    A_i[q^2_{\rm min},q^2_{\rm max}] \equiv \frac{\int_{q^2_{\rm min}}^{q^2_{\rm max}} \frac{d \left(J_i - \bar J_i \right)}{dq^2} dq^2}{\int_{q^2_{\rm min}}^{q^2_{\rm max}} \frac{d \left(\Gamma +  \bar\Gamma \right)}{dq^2}dq^2} \ .
\end{equation}
As before, only seven combinations are independent from one another. Here we consider $i=3,4,5,6s,7,8,9$. We note that here we do not consider $A_{1c}$, which has not been measured yet, and would be the asymmetric counterpart of $F_L$. Therefore, we consider $A_8$, for which measurements exist \cite{LHCb:2015svh}, instead.

Finally, we consider later also the following differential CP asymmetry:
\begin{equation}
    \frac{dA_i}{dq^2} \equiv \frac{\frac{ d\left( J_i - \bar J_i \right)}{dq^2}}{d \frac{\left( \Gamma +  \bar\Gamma \right) }{dq^2}} \ .
\end{equation}

\section{Experimental Data and Theoretical Predictions}\label{sec:data-theory}
\subsection{CKM inputs and branching ratio predictions}
The theoretical prediction for the branching ratio depends quadratically on the CKM matrix elements $|V_{ts}V_{tb}^*|$. Using CKM unitarity gives
(see \cite{DeBruyn:2022zhw}): 
\begin{equation}\label{eq:vtsvtb}
    V_{ts}V_{tb}^* = -V_{cb} \bigg[1 - \frac{\lambda^2}{2}(1-2 \bar{\rho} + 2i \bar{\eta}) \bigg] + \mathcal{O}(\lambda^6)  \ ,
\end{equation}
where $\rho$ and $\eta$ determine the apex of the Unitarity Triangle (UT). Including also the $\mathcal{O}(\lambda^2)$ terms, we obtain a tiny imaginary part for the matrix elements. 

From Eq.~\ref{eq:vtsvtb} it becomes clear that the discrepancies between the inclusive and exclusive determinations of $|V_{cb}|$ and to some extent also in $|V_{ub}|$ directly influence the determination of $V_{ts}V_{tb}^*$ \cite{DeBruyn:2022zhw} and the predictions of branching ratios for $b\to s\ell^+\ell^-$ decays as already highlighted in \cite{Fleischer:2022klb,Fleischer:2024fkm}.

To emphasize these differences, we provide SM predictions using the exclusive $|V_{cb}|$ determination and a hybrid scenario employing the exclusive $|V_{ub}|$ and inclusive $|V_{cb}|$ values \cite{DeBruyn:2022zhw}. At the order we are considering, the effect of $|V_{ub}|$ can be neglected, and the latter coincides with the inclusive scenario. For consistency, we label this the ``incl/hybrid'' scenario. 

We find 
\begin{equation}\label{eq:hybrid}
|V_{ts}V_{tb}^*|_{\rm incl/hybrid} = (41.4 \pm 0.5)\times 10^{-3} \ ,
\end{equation}
where we employ $|V_{cb}|$ from \cite{Bordone:2021oof} which agrees with the recent determination in \cite{Bernlochner:2022ucr, Finauri:2023kte}.
For the exclusive scenario, we find 
\begin{equation}\label{eq:excl}
|V_{ts}V_{tb}^*|_{\rm excl} = (38.4 \pm 0.5)\times 10^{-3},
\end{equation}
which uses the exclusive value of $|V_{cb}|$ from HFLAV \cite{HFLAV:2019otj} (see also \cite{DeBruyn:2022zhw}). The difference between \eqref{eq:hybrid} and \eqref{eq:excl} shows the long-standing $|V_{cb}|$ puzzle \cite{Gambino:2019sif,Gambino:2020jvv, Bordone:2019vic}. 

Using the inputs described above and in the previous section, we then find the following predictions in the  $1.1 \;{\rm GeV}^2 <q^2 < 6.0 \; {\rm GeV}^2$ range:
\begin{equation}\label{eq:branching_ratios_btokstarmumu}
\begin{aligned}
    \mathcal{B}(B_d \to K^{\ast 0} \mu^+\mu^-)[1.1,6.0]_{\rm incl} &= (5.50 \pm 0.70) \times 10^{-8} \ ,\\
    \mathcal{B}(B_d \to K^{\ast 0} \mu^+\mu^-)[1.1,6.0]_{\rm excl} &= (4.74 \pm 0.60) \times 10^{-8} \ .
\end{aligned}
\end{equation}
Here we do not include the uncertainty from the specific choice of long-distance model, nor do we include uncertainties on the parameters that enter the LD model. The quoted uncertainty is dominated by the uncertainty on the local form factors. We note that these values are about $10\%$ higher than those quoted in our previous work \cite{Fleischer:2024fkm}, where we used a perturbative QCD model for the long-distance effects. This highlights again the importance of pinning down the long-distance effects. At the same time, we also observe that the inclusive and exclusive CKM values introduce a larger shift in the SM prediction than the variation induced by the two long-distance models. In the remainder of this paper, we will use the inclusive CKM values unless otherwise specified.

Comparing to the measurement from the LHCb collaboration \cite{LHCb:2016ykl}
\begin{equation}\label{eq:br_btokstarmumu_exp}
     \mathcal{B}(B_d \to K^{\ast 0} \mu^+\mu^-)[1.1,6.0] = (3.42 \pm 0.30) \times 10^{-8} \ ,
\end{equation}
we find a $2.4\sigma$ deviation for the inclusive predictions and $1.9\sigma$ for the exclusive case.

Besides the clear need to resolve the inclusive versus exclusive $|V_{cb}|$ puzzle, the discrepancy between the SM and measurements shows the well-known anomalies in the $b\to s\ell^+\ell^-$ modes. Similar deviations were found in the $B\to K \ell^+\ell^-$ branching ratios (see e.g.~\cite{Fleischer:2022klb}) as well as in the vector-mode $B_s \to \phi\mu^+\mu^-$ \cite{LHCb:2021zwz,Straub:2018kue,Gubernari:2022hxn, Fleischer:2024fkm}. Here, we focus on the $B_d \to K^{\ast 0} \mu^+\mu^-$ channel, and briefly comment later on the $B_d^0\to K^{*\pm}\mu^+\mu^-$ and $B_s^0\to \phi \mu^+\mu^-$ transitions.

\subsection{Angular observables}
For the angular observables $P_i^{(\prime)}$ and $F_L$, the CKM factors drop out. Furthermore, their dependences on the local form factors are suppressed relative to the branching ratio and to the angular observables $S_i$ \cite{Descotes-Genon:2012isb}. In Table~\ref{tab:exp_constraints_cpavg} we give the measured values for these observables obtained by the LHCb collaboration \cite{LHCb:2020lmf}\footnote{A recent analysis has been performed by the CMS collaboration using different $q^2$ bins \cite{CMS-PAS-BPH-21-002,CMS:2024atz}.}. Of these, especially the  angular observable $P_5^\prime$ has received a lot of attention
\cite{ATLAS:2018gqc,LHCb:2020lmf,CMS-PAS-BPH-21-002,CMS:2024atz,Gubernari:2022hxn}, due to the large tension between its SM prediction and the measurement.

In our analysis, we take $F_L, P_1, P_2, P_4^\prime$ and $P_5^\prime$ in the bin of $[1.1,6.0]$ GeV$^2$, where hadronic long-distance effects are small. For the observables $P_3$ and $P_6^\prime$, however, we choose to consider the bin of $[6,8]$ GeV$^2$, as it is the bin nearest to the $J/\psi$ resonance that was measured in \cite{LHCb:2020lmf}. We select this bin because $P_3$ and $P_6^\prime$ are proportional to the imaginary parts of the hadronic long-distance functions, which grow largest near the $c \bar c$ resonances. For this reason, $P_3$ and $P_6^\prime$ are enhanced near the resonances.

\begin{table}[t]
    \centering
    \begin{tabular}{ll}
    Observable & Exp. value\\
    \hline
    $F_L(B_d \to K^{\ast 0}\mu^+\mu^-)[1.1,6.0]$ & $0.70 \pm 0.03$\\
    $P_1[1.1,6.0]$  & $-0.079 \pm 0.160$ \\
    $P_2[1.1,6.0]$  & $-0.162 \pm 0.051$ \\    
    $P_3[6,8]$  & $0.057 \pm 0.149$\\    
    $P_4^\prime[1.1,6.0]$  & $-0.298 \pm 0.088$ \\    
    $P_5^\prime[1.1,6.0]$  & $-0.114 \pm 0.073 $ \\
    $P_6^\prime[6,8]$  & $-0.095 \pm 0.135$  \\
    \end{tabular}
    \caption{Overview of current measurements of CP-averaged angular observables in the $B_d \to K^{\ast 0}\mu^+\mu^-$ decay by the LHCb collaboration \cite{LHCb:2020lmf}.}
    \label{tab:exp_constraints_cpavg}
\end{table}
To obtain a full picture of possible NP contributions, which may violate CP symmetry, it is important to also include CP asymmetries. 
First results on the CP asymmetries in the $B_d \to K^\ast \mu^+\mu^-$ channel have been obtained by the LHCb collaboration \cite{LHCb:2014mit,LHCb:2015svh}. We list these measurements in Table~\ref{tab:exp_constraints_cpasymm}. In the SM, all these CP-violating observables are tiny, as they are generated by the doubly-Cabibbo suppressed $\mathcal{O}^u$ operators in \eqref{eq:ham} (see \cite{Bobeth:2008ij} for detailed NLO SM predictions), which we neglect in our current work. A measurement of these CP observables would, therefore, be a sign of new CP-violating couplings interfering with strong phases from the non-local form factors $\mathcal{H}_i$. To distinguish which new coupling(s) are present, it is important to note the dependence of the observables on the different transversity form factors $\mathcal{H}_i$, which we list in Table~\ref{tab:exp_constraints_cpasymm}.

In our analysis, we use the asymmetries $\mathcal{A}_{\rm CP}^{\rm dir}, A_3, A_4, A_5$, and $A_{6s}$ in the $[6,8]$ GeV$^2$ bin, i.e. close to the $J/\psi$ resonance. In this region, the strong phase is large and the asymmetries are enhanced\footnote{A similar enhancement was pointed out for the direct CP asymmetry of the $B \to K\mu^+\mu^-$ decay in \cite{Becirevic:2020ssj, Fleischer:2022klb}} . We note that measurements even closer to the $J/\psi$ resonance, e.g. in the $[8,9]$ GeV$^2$ bin, would be expected to be even further enhanced. However, the situation is different for the $T$-odd $A_7, A_8$ and $A_9$ asymmetries, which contain terms that are proportional to either the real parts of long-distance parameters \cite{Bobeth:2008ij}, and/or independent of the long-distance effects. The latter is indicated in Table~\ref{tab:exp_constraints_cpasymm} by a $1$. Therefore, these observables are predicted to be smaller close to the resonance region. We, therefore, consider these $T$-odd observables in $q^2 \in [1.1,6.0]$ GeV$^2$ region, away from the $c \bar c$ resonance.  Alternatively, these asymmetries are also very sensitive probes of NP effects in the high-$q^2$ region beyond the $\psi(2S)$ resonance, but this requires going beyond the long-distance model of \cite{Bobeth:2017vxj}. 

We note that for the CP asymmetries $A_7, A_8$, and $A_9$ we have increased the uncertainties reported in \cite{LHCb:2015svh} to be more conservative and avoid averaging over $q^2$ bins, which could wash out resonance effects. Specifically, we assume uncertainties that approximately account for the variance across the bins in the $[1.1,6.0]$ GeV$^2$ $q^2$ region.

\begin{table}[]
    \centering
    \begin{tabular}{lll}
Observable & Exp. value & Proportional to\\
\hline
    $\mathcal{A}_{\rm CP}^{\rm dir}(B_d\to K^{\ast 0} \mu^+ \mu^-)[7,8]$  & $0.099 \pm 0.087$ \cite{LHCb:2014mit}  & $\Im \mathcal{H}_\perp, \Im \mathcal{H}_\parallel, \Im \mathcal{H}_0$\\
    $A_3[6,8]$ & $0.064 \pm 0.068$ \cite{LHCb:2015svh}  & $\Im \mathcal{H}_\perp, \Im \mathcal{H}_\parallel$\\
    $A_4[6,8]$ & $-0.037 \pm 0.074$ \cite{LHCb:2015svh} & $\phantom{\Im \mathcal{H}_\perp,} \Im \mathcal{H}_\parallel, \Im \mathcal{H}_0$\\
    $A_5[6,8]$ & $0.129 \pm 0.068$ \cite{LHCb:2015svh} & $\Im \mathcal{H}_\perp, \phantom{\Im \mathcal{H}_\parallel,} \Im \mathcal{H}_0$\\
    $A_{6s}[6,8]$ & $0.047 \pm 0.063$ \cite{LHCb:2015svh} & $\Im \mathcal{H}_\perp, \Im \mathcal{H}_\parallel$ \\
\hline
    $A_7[1.1,6.0]$ & $-0.045 \pm 0.100^\ast$ \cite{LHCb:2015svh} & $\phantom{\phantom{\Re \mathcal{H}_\perp,}} \Re \mathcal{H}_\parallel, \Re \mathcal{H}_0, 1$\\
    $A_8[1.1,6.0]$ & $-0.047 \pm 0.100^\ast$ \cite{LHCb:2015svh} & $\Re \mathcal{H}_\perp, \phantom{\Re \mathcal{H}_\parallel,} \Re \mathcal{H}_0, 1$\\
    $A_9[1.1,6.0]$ & $-0.033 \pm 0.200^\ast$ \cite{LHCb:2015svh} & $ \Re \mathcal{H}_\perp, \Re \mathcal{H}_\parallel, \phantom{\Re \mathcal{H}_0, \, } 1$\\
    \end{tabular}
    \caption{Measurements of CP asymmetries in the $B_d \to K^{\ast 0}\mu^+\mu^-$ decay, along with their dependences on the hadronic long-distance functions $\mathcal{H}_{\perp, \parallel, 0}$. Observables containing terms which do not depend on any of the long-distance functions are marked as proportional to $1$.
    The uncertainties marked by asterisks are inflated as discussed in the text.}
    \label{tab:exp_constraints_cpasymm}
\end{table}

\section{Sensitivity of the Observables to NP}\label{sec:sens-obs}
Measurements of CP-averaged observables in the $B_d \to K^\ast \mu^+\mu^-$ decay have previously been used to determine the Wilson coefficients $\cplus_9$, $\cplus_{10}$, $\cminus_9$ and $\cminus_{10}$ under the assumption that these coefficients are real \cite{LHCb:2024onj}. Our goal is now to determine the complex values of these coefficients. To this end, we utilize observables which allow for theoretically clean extractions of the coefficients in question.

As stated in the introduction, the $B\to K\ell^+\ell^-$ modes depend {\it exclusively} on $\cplus_9$ and $\cplus_{10}$. This feature has the important consequence that their (complex) values can be extracted from branching ratios and CP measurements in $B \to K\mu^+\mu^-$ decays, following the strategy discussed in \cite{Fleischer:2022klb}. This extraction is clean in the sense that possible scalar and pseudoscalar NP interactions were shown to be negligible in these semileptonic decays \cite{Fleischer:2022klb}. 
 
Consequently, we may focus our analysis of $B \to K^\ast \mu^+\mu^-$ on extracting $\cminus_9$ and $\cminus_{10}$.  The analysis of \cite{Fleischer:2022klb} gives a range of allowed complex values for $\cplus_9$ and $\cplus_{10}$ within the currently available data. To illustrate our strategy we fix $\cplus_9$ and $\cplus_{10}$ to the values of Benchmark Point 2 in \cite{Fleischer:2022klb}:
\begin{equation}\label{eq:c9plus_c10plus_bm}
\begin{aligned}
\mathcal{C}_9^{+ \rm NP} &= 0.30 \abs{C_9^{\rm SM}} e^{i 220^\circ} = -0.98 - 0.82 i  \ ,\\
\mathcal{C}_{10}^{+ \rm NP} &= -\mathcal{C}_9^{+ \rm NP} \ .
\end{aligned}
\end{equation}
However, we have also explicitly checked that our analysis and conclusions also hold for other benchmark points allowed by the $B\to K \ell^+ \ell^-$ data.

\begin{figure}
    \centering
    \includegraphics[width=0.75\linewidth]{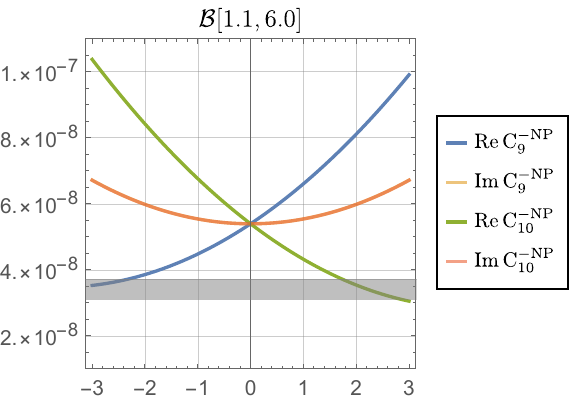}
    \caption{Sensitivity of the branching ratio of $B_d \to K^\ast \mu^+\mu^-$ to NP contributions to the real and imaginary parts of $\mathcal{C}_9^{- \rm NP}$ and $\mathcal{C}_{10}^{- \rm NP}$, while $\cplus_9$ and $\cplus_{10}$ are fixed to the values in \eqref{eq:c9plus_c10plus_bm}. The horizontal band shows the measurement from \cite{LHCb:2016ykl}.}
    \label{fig:obs_dep_br}
\end{figure}

Our goal is to identify observables which exhibit strong dependences on $\mathcal{C}_9^{- \rm NP}$ and $\mathcal{C}_{10}^{- \rm NP}$. To this end, we determine how the observables vary as functions of the Wilson coefficients. In Fig.~\ref{fig:obs_dep_br}, we show the branching ratio of $B_d^0 \to K^{0\ast} \mu^+\mu^-$ as a function of the real and imaginary parts of $\mathcal{C}_9^{- \rm NP}$ and $\mathcal{C}_{10}^{- \rm NP}$. Here, we set $\mathcal{C}_9^{+ \rm NP}$ and $\mathcal{C}_{10}^{+ \rm NP}$ to the values in \eqref{eq:c9plus_c10plus_bm}, and include interference between the $\mathcal{C}_+$ and $\mathcal{C}_-$ coefficients. 
The gray bar indicates the LHCb measurement \cite{LHCb:2016ykl} also given in \eqref{eq:br_btokstarmumu_exp}. We observe that to accommodate the data, we need NP contributions to the real part of $\mathcal{C}_9^{- \rm NP}$, and/or of $\mathcal{C}_{10}^{- \rm NP}$. We note that the dependence on the imaginary parts of $\mathcal{C}_9^{- \rm NP}$  (yellow) and $\mathcal{C}_{10}^{- \rm NP}$ (red) is overlapping at this scale. In Fig.~\ref{fig:obs_dep_br}, we only consider the dependence on one Wilson coefficient, setting the others to zero. As such, the figure does not include interference terms which involve both $\mathcal{C}_9^{- \rm NP}$ and $\mathcal{C}_{10}^{- \rm NP}$. We do not include an uncertainty on the NP predictions because Fig.~\ref{fig:obs_dep_br} is merely to study the dependence of the branching ratio on the Wilson coefficients. However, as discussed above, the branching ratio measurement is in tension with the SM prediction, hinting at NP in $\cminus_9$ and/or $\cminus_{10}$. In order to disentangle what type of NP is required, it is important to consider additional observables. 

In Fig.~\ref{fig:observ_dep_bm_point} and in Appendix~\ref{sec:sensNP}, we show the sensitivity to the NP Wilson coefficients of the observables. 
We observe that $P_3$, $P_6^\prime$, $A_3$, $A_4$, $A_5$, and $A_{6s}$ change minimally, even with large NP contributions. For this reason, they are not efficient for isolating NP contributions to specific coefficients. Other observables are more promising and stand out in their sensitivity to the real or imaginary part of a specific coefficient. In fact, we can find a minimal set of four observables which can isolate the four NP contributions;
\begin{itemize}
    \item The forward-backward asymmetry $F_L$ is highly sensitive to real NP contributions to $\cminus_9$.
    \item The direct CP asymmetry $\mathcal{A}_{\rm CP}^{\rm dir}$ is nonzero only if there exists an imaginary part in $\cminus_9$.
    \item The CP asymmetry $A_7$ is always zero if the imaginary part of $\cminus_{10}$ is zero. 
    \item $P_1$ gives access to the real part of $\cminus_{10}$.
    \end{itemize}

In the next section, we demonstrate that using these four observables alone allows us to disentangle the possible NP coefficients. We note that the branching ratio and $P_5^\prime$ are also good probes of NP couplings. However, since they are not sensitive to one dominant NP coupling, we do not consider them in our minimal set. A full global analysis would obviously include these observables. In our strategy, they can also be added and they allow for a consistency check of the data including NP. More interestingly, they may be also used to disentangle NP and hadronic effects following a similar method as in \cite{Fleischer:2022klb}. To summarize, we collect the dependences of the various observables on $\cminus_9$ and $\cminus_{10}$ in Table~\ref{tab:obs_dep}, where a checkmark indicates dependence, two checkmarks indicate strong dependence, and no checkmark indicates no or negligible dependence.
\begin{figure}
    \centering
    \subfloat{\includegraphics[width=0.35\linewidth]{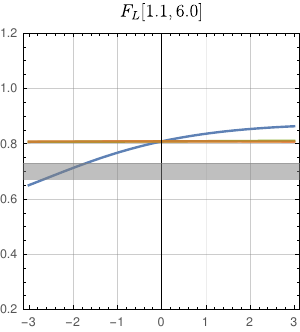}}
\hspace{1.3cm}
    \subfloat{\includegraphics[width=0.355\linewidth]{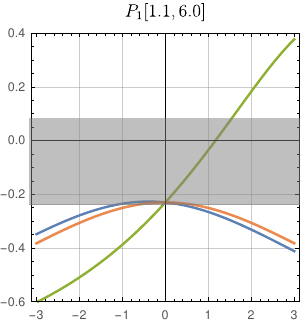}}\\
        \vspace{0.4cm}
        \subfloat{\includegraphics[width=0.3655\linewidth]{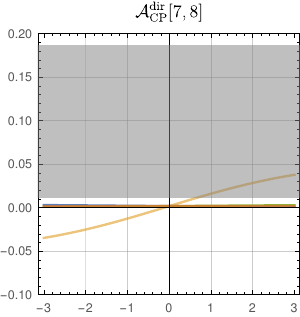}}
       \hspace{1.3cm}
         \subfloat{\includegraphics[width=0.35\linewidth]{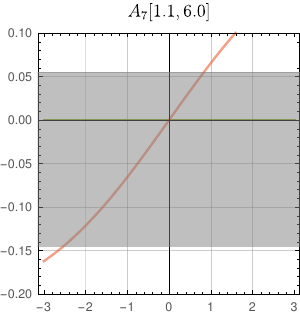}} \\   \vspace{0.4cm}      \subfloat{\includegraphics[width=0.45\linewidth]{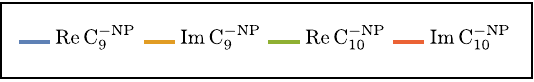}} 
    \caption{Sensitivity of our nominal set of four observables on the variation of the complex Wilson coefficients. Here we set $\cplus_9$ and $\cplus_{10}$ to the values in \eqref{eq:c9plus_c10plus_bm} and then turn on the real and imaginary parts of $\mathcal{C}_9^{- \rm NP}$ and $\mathcal{C}_{10}^{- \rm NP}$ independently. The grey bars indicate the experimental measurements in Table~\ref{tab:exp_constraints_cpavg}.}
    \label{fig:observ_dep_bm_point}
\end{figure}

\begin{table}[]
    \centering
    \begin{tabular}{l|llll}
    Observable     & $\Re \cminus_9$ & $\Im \cminus_9$ & $\Re \cminus_{10}$ & $\Im \cminus_{10}$ \\
    $\mathcal{B}[1.1,6.0]$ & $\checkmark \checkmark$ & $\checkmark$& $\checkmark \checkmark$ & $\checkmark$ \\
    $F_L[1.1,6.0]$ & $\checkmark \checkmark$ & &  & \\
    $P_1[1.1,6.0]$ &$\checkmark$& $\checkmark$ & $\checkmark \checkmark$ & $\checkmark$\\ 
    $P_2[1.1,6.0]$ &$\checkmark \checkmark $& $\checkmark$&$\checkmark$ & $\checkmark$\\ 
    $P_3[6,8]$ &  &  &  & \\
    $P_4^\prime[1.1,6.0]$ & $\checkmark$ & $\checkmark$ & $\checkmark$ & $\checkmark$\\
    $P_5^\prime[1.1,6.0]$ & $\checkmark \checkmark$ & $\checkmark$ & $\checkmark \checkmark$ & $\checkmark$\\
    $P_6^\prime[6,8]$ & & & & \\
    \hline
    $\mathcal{A}_{\rm CP}^{\rm dir}[7,8]$ & & $\checkmark$ & &\\
    $A_3[6,8]$ & & & &\\
    $A_4[6,8]$ & & $\checkmark$ & &\\
    $A_5[6,8]$ & & & &\\
    $A_{6s}[6,8]$ & & & &\\
    $A_7[1.1,6.0]$ & & & & $\checkmark \checkmark$\\
    $A_8[1.1,6.0]$ &$\checkmark$ & $\checkmark$& $\checkmark$& $\checkmark$\\
    $A_9[1.1,6.0]$ & & & & $\checkmark$\\
    \end{tabular}
    \caption{Summary of the sensitivity of observables of the $B_d \to K^\ast \mu^+\mu^-$ decay to the variation of complex Wilson coefficients. A checkmark indicates dependence, two checkmarks indicate strong dependence, and no checkmark indicates no or negligible dependence.}
    \label{tab:obs_dep}
\end{table}

In the above, we focused on the $B_d^0\to K^{*0}\mu^+\mu^-$ decay, for which several measurements are available. Similar information on the Wilson coefficients could be obtained from the charged $B^\pm \to K^{\ast \pm} \mu^+\mu^-$ decay. On the other hand, the $B_s \to \phi\mu^+\mu^-$ channel, which only differs from the $B_d^0 \to K^{\ast 0} \mu^+\mu^-$ decay through the spectator quark, offers interesting new observables due to the sizeable $B_s^0$ meson decay width difference $\Delta\Gamma_s$. Measuring the time-dependent angular coefficients gives access to mixing-induced angular CP asymmetries $s_i$ and $h_i$ (see \cite{Descotes-Genon:2015hea}) which, to the best of our knowledge, have not yet been measured. For completeness, we study the dependencies of these observables on the complex Wilson coefficients in Appendix~\ref{sec:timedep}. We note that none of these observables show larger sensitivity to the NP couplings than the direct CP asymmetries.

\section{Extracting the Complex Values of $\cminus_9$ and $\cminus_{10}$}\label{sec:sec-extr}

In this section, we show how measurements of a minimal set of four observables in $B_d^0 \to K^{\ast 0} \mu^+\mu^-$ let us transparently extract the complex values of $\cminus_9$ and $\cminus_{10}$, using input on $\cplus_9$ and $\cplus_{10}$ from $B \to K\mu^+\mu^-$ \cite{Fleischer:2022klb,Fleischer:2024fkm}. Moreover, we use
\begin{itemize}
    \item $F_L$ in the $q^2$ bin of $[1.1,6.0]$ GeV$^2$, which probes $\rm{Re}\;\cminus_9$.
    \item $P_1$ in the $q^2$ bin of $[1.1,6.0]$ GeV$^2$, which probes $\rm{Re}\;\cminus_{10}$.
    \item $A_7$ in the $q^2$ bin of $[1.1,6.0]$ GeV$^2$, which probes $\rm{Im}\;\cminus_{10}$.
    \item $\mathcal{A}_{\rm CP}^{\rm dir}$ in the $q^2$ bin of $[7,8]$ GeV$^2$, which probes $\rm{Im}\;\cminus_{9}$.
\end{itemize}

\subsection{Illustration using current data}
\begin{figure}
    \centering
    \subfloat[Re-Im plane]
       { \includegraphics[width=0.45\linewidth]{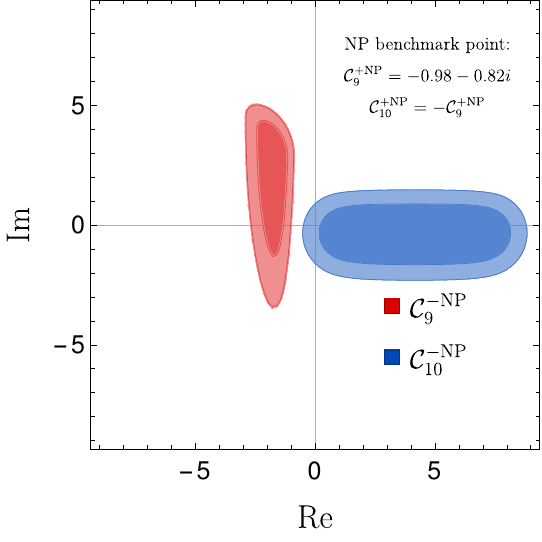}}
    \hfill
    \subfloat[Arg-Abs plane]{\includegraphics[width=0.45\linewidth]{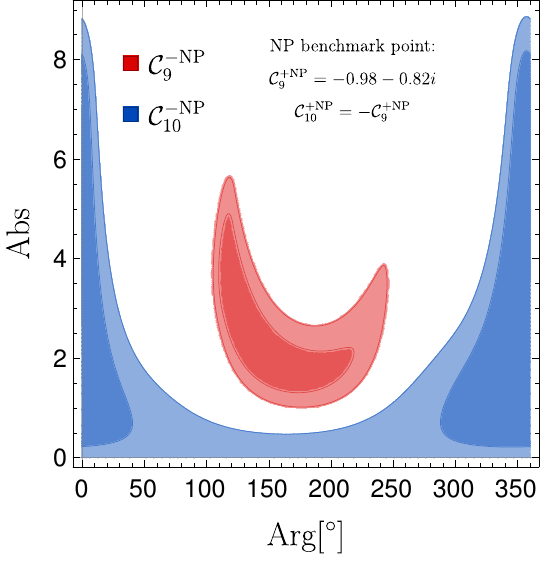}}
    \caption{Constraints on $\mathcal{C}_9^{- \rm NP}$ and $\mathcal{C}_{10}^{- \rm NP}$ in the complex plane, obtained from $F_L$, $\mathcal{A}_{\rm CP}^{\rm dir}[7,8]$, $P_1$ and $A_7$ using current experimental data and assuming the $\cplus_{9}$, $\cplus_{10}$ benchmark point given in \eqref{eq:c9plus_c10plus_bm}. The contours show the $68 \, \%$ and $90 \, \%$ confidence intervals.}
    \label{fig:fitplot}
\end{figure}

Let us now illustrate how measurements of $F_L$, $\mathcal{A}_{\rm CP}^{\rm dir}$, $P_1$ and $A_7$ complement each other and allow us to determine the complex values of $\cminus_9$ and $\cminus_{10}$. In Fig.~\ref{fig:fitplot}, we show the constraints on $\mathcal{C}_9^{- \rm NP}$ and $\mathcal{C}_{10}^{- \rm NP}$ from current data on the observables provided in Tables~\ref{tab:exp_constraints_cpavg} and~\ref{tab:exp_constraints_cpasymm}. We assume for $\cplus_9$ and $\cplus_{10}$ the values in \eqref{eq:c9plus_c10plus_bm}. We note that those Wilson coefficients are in agreement with the current data from $B\to K\mu^+\mu^-$ decays, but they present a benchmark scenario. To illustrate the complementarity of observables, we neglect theoretical uncertainties. As can be seen in Fig.~\ref{fig:fitplot}, our analysis indicates  $\Re \mathcal{C}_9^{-\rm NP}<0$, which is driven by $F_L$.  

At the same time, this analysis also leaves large room for CP-violating NP contributions: the $68 \%$ confidence intervals give roughly $\Re \mathcal{C}_9^{-\rm NP} \in [-2.5, 1]$, $\Im \mathcal{C}_9^{-\rm NP} \in [-1, 4]$, $\Re \mathcal{C}_{10}^{-\rm NP} \in [0, 8]$, and $\Im \mathcal{C}_{10}^{-\rm NP} \in [-2, 1]$. The large allowed regions for all parameters expect $\Re \mathcal{C}_9^{- \rm NP}$ can be understood from the current experimental precision: while $F_L$ requires $\Re \mathcal{C}_9^{-\rm NP}<0$, the other observables have very large uncertainties, e.g.~$200 \, \%$ in the case of $P_1$. With increased precision, the regions will shrink, thereby allowing for a transparent determination of the complex $\mathcal{C}_9^{-\rm NP}$ and $\mathcal{C}_{10}^{-\rm NP}$. 

\subsection{Future Scenario}
\begin{figure}
    \centering
    \hfill
    \subfloat{\includegraphics[height=0.35\linewidth]{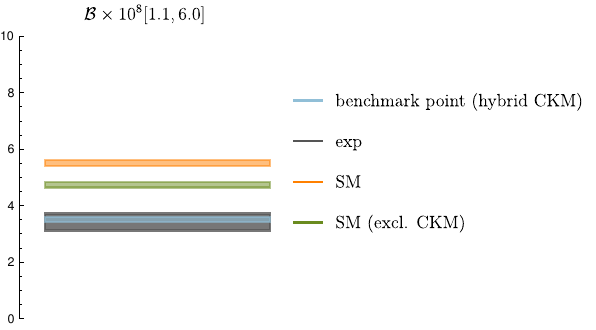}}\\
    \hfill
    \subfloat{\includegraphics[height=0.3\linewidth]{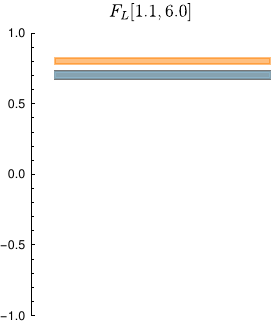}}
    \hfill
    \subfloat{\includegraphics[height=0.3\linewidth]{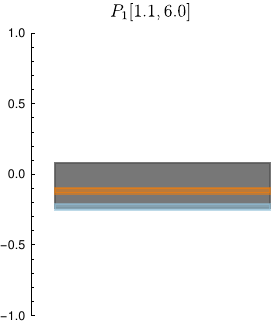}}
    \hfill
    \subfloat{\includegraphics[height=0.3\linewidth]{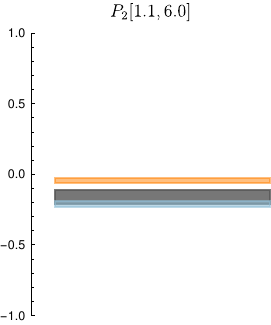}}\\
    \hfill\\
    \hfill
    \subfloat{\includegraphics[height=0.3\linewidth]{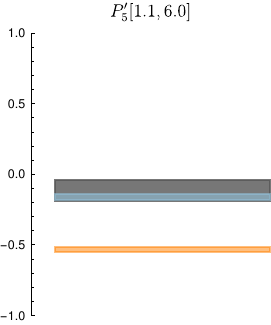}}
    \hfill
   \subfloat{\includegraphics[height=0.3\linewidth]{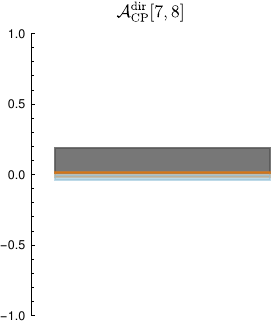}}
    \hfill
    \subfloat{\includegraphics[height=0.3\linewidth]{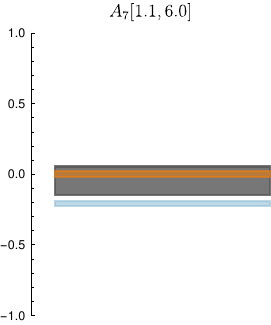}}
    \caption{Illustration of the predictions at the benchmark point in \eqref{eq:c9minus_c10minus_bm} (blue) compared to their SM predictions (orange) and the experimental data (gray) given in Tables~\ref{tab:exp_constraints_cpavg} and \ref{tab:exp_constraints_cpasymm}. For the branching ratio, we give the SM prediction using the inclusive (orange) and exclusive (green) CKM parameters from \eqref{eq:branching_ratios_btokstarmumu}}
    \label{fig:agreement_plots}
\end{figure}

In order to illustrate the prospects, we consider a future scenario with reduced uncertainties for the four observables. For this, we assume:
\begin{equation}\label{eq:c9minus_c10minus_bm}
\begin{aligned}
\mathcal{C}_9^{- \rm NP} &= -2.00 -1.00 i = \ 2.24 e^{-i 153^\circ },\\
\mathcal{C}_{10}^{- \rm NP} &= 1.00 - 2.00 i =  2.24 e^{-i 63^\circ}\ ,\\
\end{aligned}
\end{equation}
and the benchmark for $\cplus_{9,10}$ in \eqref{eq:c9plus_c10plus_bm}. Specifically, combining both expressions, we have as a benchmark:
\begin{equation}
\begin{aligned}
C_9^{\rm NP} &= -1.49 - 0.91 i = 1.75 e^{-i 149^\circ}\\
C_9^{\prime \rm NP} &= 0.51 + 0.09 i = 0.52 e^{i 10^\circ}\\
C_{10}^{\rm NP} &= 0.99 - 0.59 i = 1.15 e^{-i 31^\circ}\\
C_{10}^{\prime \rm NP} &= -0.01 + 1.41 i = 1.41 e^{i 90^\circ}\ .
\end{aligned}
\end{equation}
We have chosen this point to feature complex Wilson coefficients while remaining consistent with current data, which we illustrate in Fig.~\ref{fig:agreement_plots}. In this illustration, we show our observables, the branching ratio, $P_5'$ and $P_2$ at the benchmark point \eqref{eq:c9minus_c10minus_bm} (blue), compared to the data (gray) and to the SM predictions (orange). We include the absolute experimental uncertainty, however, we do not include uncertainties on the predictions. We find that our benchmark NP scenario accommodates the data on $F_L$, $P_5'$ and the branching ratio. Also for the other observables, which have large experimental uncertainties, and our benchmark point agrees with the data. As expected, we find that these observables are rather insensitive to NP and as such do not significantly change with respect to their SM prediction when allowing for NP contributions. 

\begin{figure}
    \centering
    \subfloat[Re-Im plane]{
    \includegraphics[width=0.45\linewidth]{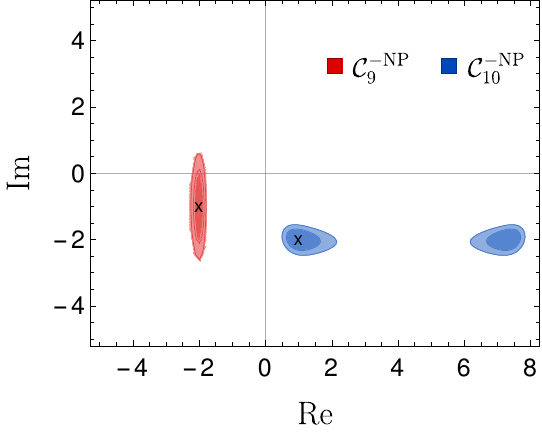}\label{fig:fitplot_complexbm_a}}
    \hfill
    \subfloat[Arg-Abs plane]{
    \includegraphics[width=0.45\linewidth]{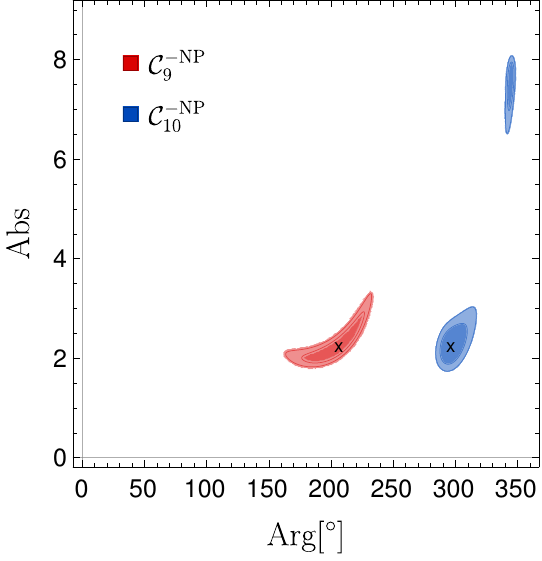}\label{fig:fitplot_complexbm_b}}
    \caption{Constraints on $\mathcal{C}_9^{- \rm NP}$ and $\mathcal{C}_{10}^{- \rm NP}$ in the complex plane in our future scenario. The crosses indicate the benchmark point given in \eqref{eq:c9minus_c10minus_bm}.}
    \label{fig:fitplot_allobs}
\end{figure}

Specifically, for our benchmark point in \eqref{eq:c9minus_c10minus_bm}, we find
\begin{equation}\label{eq:bm_obs_values}
\begin{aligned}
    F_L[1.1,6.0] &= 0.70 \pm 0.006 \ ,\\
    \mathcal{A}_{\rm CP}^{\rm dir}[7,8] &= -0.02 \pm 0.02 \ ,\\
    P_1[1.1,6.0] &= -0.23 \pm 0.03 \ ,\\
    A_7[1.1,6.0] &= -0.21 \pm 0.02 \ .
\end{aligned}
\end{equation}
Here we assume an uncertainty a factor five smaller than the current experimental precision.   
In Fig.~\ref{fig:fitplot_allobs}, we show how these values translate into constraints on the complex $\mathcal{C}_{9, 10}^{- \rm NP}$ coefficients. The crosses indicate the values of \eqref{eq:c9minus_c10minus_bm}. We see that in this scenario with reduced uncertainties we indeed gain a sharper picture of the complex coefficients. In particular, even with a significant complex phase in $\cminus_{10}$, the observables constrain the coefficients efficiently.

Consequently, we see that precise measurements of only four observables in $B_d \to K^{\ast 0} \mu^+\mu^-$ (or in either of the partner decays $B^\pm \to K^{\ast \pm} \mu^+\mu^-$ and $B_s \to \phi\mu^+\mu^-$), allow us to determine the complex coefficients $\cminus_9$ and $\cminus_{10}$. Combining this information with the analysis of $B \to K\mu^+\mu^-$, which gives access to $\cplus_9$ and $\cplus_{10}$, it is possible to extract the complex coefficients $C_9^{(\prime)}$ and $C_{10}^{(\prime)}$. Finally, we may perform a simultaneous analysis of the $B \to K\mu^+\mu^-$ and $B \to K^\ast \mu^+\mu^-$ systems, similar to the angular analyses in \cite{LHCb:2023gpo} and \cite{LHCb:2016due}.

\begin{figure}[t]
    \centering
    \subfloat{\includegraphics[width=0.35\textwidth]{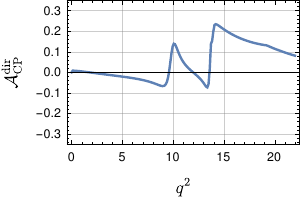}}
    \hspace{2cm}
    \subfloat{\includegraphics[width=0.35\textwidth]{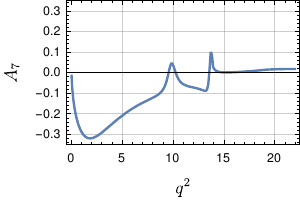}}\\
    \subfloat{\includegraphics[width=0.35\textwidth]{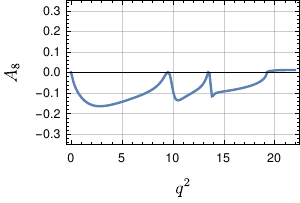}}
    \hspace{2cm}
    \subfloat{\includegraphics[width=0.35\textwidth]{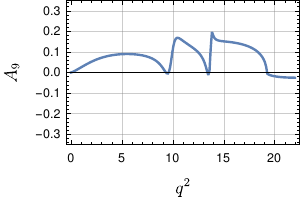}}
    \caption{Differential CP asymmetries evaluated at the benchmark point given in \eqref{eq:c9minus_c10minus_bm}.}
    \label{fig:dACPplots}
\end{figure}

We note that, if indeed $\cminus_{10}$ is sizeable as in our benchmark, the differential CP asymmetries $\mathcal{A}_{\rm CP}^{\rm dir}$ and $A_{7,8,9}$ become sensitive to the phase of the long-distance effects. We illustrate this in Fig.~\ref{fig:dACPplots}, where we fix $\cminus_9$ and $\cminus_{10}$ to the values in \eqref{eq:c9minus_c10minus_bm}. We find that the direct CP asymmetry, which is proportional to the imaginary parts of hadronic long-distance parameters, as shown in Table~\ref{tab:exp_constraints_cpasymm}, is enhanced near the $J/\psi$ and $\psi(2S)$ resonances at $q^2 = m_{J/\psi}^2 = 9.6$ GeV$^2$ and $q^2 = m_{\psi_{2S}} = 13.6$ GeV$^2$. In contrast, the asymmetries $A_7$, $A_8$, and $A_9$, which contain terms that do not depend on any long-distance parameters, are suppressed near the resonances and instead grow large in regions of the $q^2$ spectrum further from the peaks. Notably, the CP asymmetry $A_7$ reaches $-0.3$ in the low-$q^2$ region. It would be interesting to see if future analyses could utilize this enhancement to probe hadronic long-distance effects. We stress that for the asymmetries $A_7$, $A_8$, and $A_9$ this enhancement also appears in the low-$q^2$ region.

\subsection{\boldmath Connection to leptonic $B_s\to \mu^+\mu^-$ decays}\label{sec:bsmumu}
If indeed NP appears in $C_{10}^{(\prime)}$, this would also affect the leptonic $B_s \to \mu^+\mu^-$ decay \cite{DeBruyn:2012wk,Fleischer:2017yox,Fleischer:2024fkm}. This decay is experimentally well established, with the following experimental world average \cite{ParticleDataGroup:2024cfk}:
\begin{equation}
    \overline{\mathcal{B}}(B_s \to \mu^+\mu^-)=(3.34 \pm 0.27)\times 10^{-9} \ ,
    \label{eq:Bsmumu_BR_HFLAV}
\end{equation}
where $\overline{\mathcal{B}}$ denotes the time-integrated branching ratio (see, e.g,.~\cite{Fleischer:2024fkm} for the detailed formalism). 
This measurement agrees within $1\sigma$ with the SM prediction in \cite{Fleischer:2024fkm} (see also \cite{Beneke:2019slt}). However, it is important to note that this decay also depends on possible (pseudo)-scalar NP coefficients $\cminus_P$ and $\cminus_S$. Therefore, as discussed in detail in \cite{Fleischer:2024fkm}, even though the branching ratio is in agreement with the experimental measurement, there is still sufficient room for NP coefficients when allowing for (pseudo)-scalar interactions. Yet, the $B_s^0 \to \mu^+\mu^-$ decay by itself is not sufficient to disentangle these NP interaction. However, combining this decay with the information from our analysis of $B\to K^*\mu^+\mu^-$, which determines $\cminus_{10}$, allows to convert the measurement of $B_s^0 \to \mu^+\mu^-$ into a constraint on the (pseudo)-scalar Wilson coefficients. 

To illustrate this point, we use our benchmark in \eqref{eq:c9minus_c10minus_bm} to predict the leptonic decay rate. We find
\begin{equation}
\begin{aligned}
    \overline{\mathcal{B}}(B_s \to \mu^+\mu^-) &= (2.86 \pm 0.07) \times 10^{-9} \ ,
\end{aligned}
\end{equation}
which has a $1.7 \sigma$ tension with the experimental measurement. Specifically, within $1\sigma$, we then find that $|C_S-C_S'|<0.12$ and $|C_P-C_P'|<0.21$, thus strongly limiting the allowed parameter space for these NP contributions. The possible complex (pseudo)-scalar NP could be further disentangled by future measurements of the CP asymmetries in $B_s^0 \to \mu^+\mu^-$, specifically by measurement $\mathcal{A}_{\rm CP}^{\Delta\Gamma}$ accessible through measurements of the effective lifetime and the mixing-induced CP asymmetry \cite{Fleischer:2024fkm}. First measurements of $\mathcal{A}_{\rm CP}^{\Delta\Gamma}$ are available from LHCb, CMS and Atlas, and a future tagged time-dependent analysis giving access to the mixing-induced CP asymmetry is highly anticipated.

\section{Conclusion}\label{sec:concl}
We have studied the semileptonic decays $B_{(s)} \to V \ell^+\ell^-$ with $V \in \{\phi, K^\ast\}$, considering the general case of CP-violating NP contributions. Compared with the $B \to K\mu^+\mu^-$ decays, these decays offer complementary dependences on the Wilson coefficients $C_9^{(\prime)}$ and $C_{10}^{(\prime)}$, providing additional avenues for disentangling possible NP contributions. We previously presented a method to utilise measurements of the branching ratios and CP asymmetries of the $B \to K\mu^+\mu^-$ modes to determine transparently the coefficients $\cplus_9 = C_9 + C_9^\prime$ and $\cplus_{10} = C_{10} + C_{10}^\prime$ which are complex in the presence of CP-violating NP contributions. On the other hand, the $B_{(s)} \to V \mu^+\mu^-$ modes depend on both $\cplus_9$ and $\cplus_{10}$ as well as on $\cminus_9$ and $\cminus_{10}$. Having both combinations available, we can determine the individual coefficients $C_9^{(\prime)}$ and $C_{10}^{(\prime)}$. In this work, we have presented and illustrated a corresponding strategy, using specifically the $B_d \to K^{\ast 0}\mu^+\mu^-$ decay. However, the method can also be applied to $B^\pm \to K^{\ast \pm} \mu^+\mu^-$ or $B_s \to \phi \mu^+\mu^-$ modes, as well as their counterparts with $e^+e^-$ in the final states.

Surveying the available observables in the $B_d \to K^{\ast 0} \mu^+\mu^-$ decay, we have identified a minimal set of four observables with strong dependences on the real or imaginary parts of $\cminus_9$ and $\cminus_{10}$. These observables are the longitudinal polarization fraction $F_L$, the angular observable $P_1$ and the direct CP asymmetries $\mathcal{A}_{\rm CP}^{\rm dir}$ and $A_7$. We have considered the asymmetry $\mathcal{A}_{\rm CP}^{\rm dir}$ in the invariant mass bin of $q^2 \in [7,8]$ GeV$^2$ and $A_7$ in $q^2 \in [1.1,6.0]$ GeV$^2$ to benefit from the enhancement of these CP asymmetries by hadronic long-distance effects near and far from the $c \bar c$ resonance regions, respectively. Using current data on these four observables, we have demonstrated how they allow a transparent determination of the complex Wilson coefficients. We find $68 \, \%$ confidence intervals of approximately $\Re \mathcal{C}_9^{- \rm NP} \in [-2.5,-1]$, $\Im \mathcal{C}_9^{- \rm NP} \in [-1.5, 4.5]$, $\Re \mathcal{C}_{10}^{- \rm NP} \in [0, 8]$, and $\Re \mathcal{C}_{10}^{- \rm NP} \in [-2, 1]$. We observe that the intervals are still quite large, encompassing the SM points for $\Im \mathcal{C}_{9}^{- \rm NP}$, $\Re \mathcal{C}_{10}^{- \rm NP}$, and $\Im \mathcal{C}_{10}^{- \rm NP}$ but also leaving significant room for possible CP-violating phases originating from physics beyond the SM. In order to illustrate how the situation may change in the future, we considered a NP benchmark scenario assuming a factor $5$ improvement in precision on all relevant observables with respect to the current picture: We found that we could then determine the real and imaginary parts of the Wilson coefficients with $68 \, \%$ confidence intervals of approximately $\Re \mathcal{C}_9^{- \rm NP} \in [-2.25,-1.75]$, $\Im \mathcal{C}_9^{- \rm NP} \in  [-2, 0]$, $\Re \mathcal{C}_{10}^{- \rm NP} \in [0.5, 1.5]$, and $\Im \mathcal{C}_{10}^{- \rm NP} \in [-2.25,-1.75]$. Needless to note, the central values will depend on the actual measured values. However, the chosen scenario demonstrates the potential of these four observables and our method for extracting the CP-violating Wilson coefficients.

The $\cminus_{10}$ coefficient determined through our strategy from rare semileptonic decays is a key input for analyses of the leptonic decay $B_s \to \mu^+\mu^-$, allowing us to constrain possible (pseudo-)scalar NP contributions, as we discussed in previous work. Following these lines, we get a powerful synergy between semileptonic and leptonic rare $B$ decays in the testing of the SM and the search for NP contributions.



We look forward to improved precision on the angular observables of the semileptonic $B_{(s)} \to V \ell^+\ell^-$ with $V \in \{\phi, K^\ast\}$. Furthermore, we encourage our experimental colleagues also to refine the measurements of the CP asymmetries, which, as we have shown, are essential for pinning down possible CP-violating NP contributions.

\section*{Acknowledgments}
We are grateful to Eleftheria Malami for valuable discussions.
This research has been supported by the Netherlands Organisation for Scientific Research
(NWO). The work of K.K.V. is supported by the Dutch Research Council (NWO) as part of the project Solving Beautiful Puzzles (VI.Vidi.223.083) of the research programme Vidi. 

\newpage
\appendix
\section{Sensitivity to NP Wilson coefficients}
\label{sec:sensNP}

\begin{figure}
    \centering
    \subfloat{\includegraphics[width=0.3\linewidth]{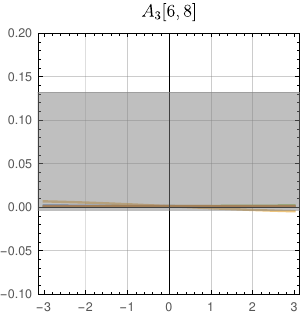}}
    \hfill
    \subfloat{\includegraphics[width=0.3\linewidth]{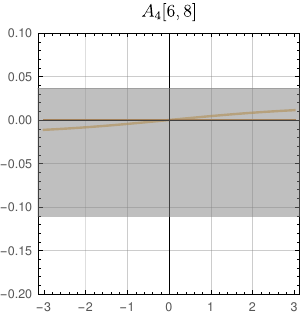}}
    \hfill
    \subfloat{\includegraphics[width=0.3\linewidth]{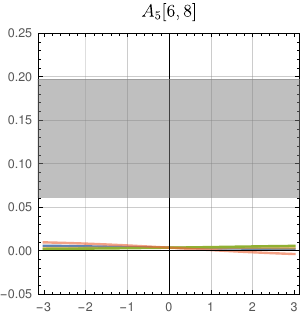}}
    \hfill\\
    \subfloat{\includegraphics[width=0.3\linewidth]{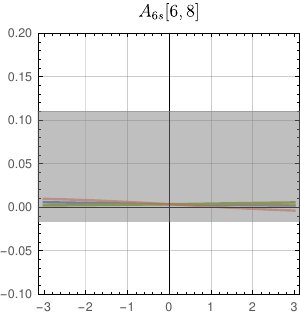}}
    \hfill
    \subfloat{\includegraphics[width=0.3\linewidth]{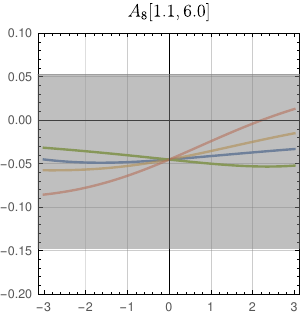}}
    \hfill
    \subfloat{\includegraphics[width=0.3\linewidth]{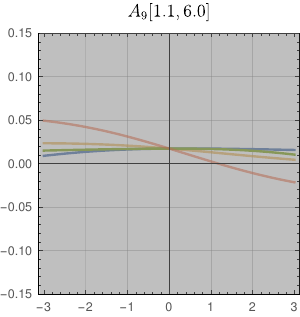}}\\
    \vspace{0.4cm}
    \subfloat{\includegraphics[width=0.3\linewidth]{Figures/obs_dep_plots_bm_point/label.pdf}}
    \hfill
    \caption{Sensitivity of the direct CP asymmetries on the variation of the complex Wilson coefficients as in Fig.~\ref{fig:observ_dep_bm_point}.}
\label{fig:observ_dep_CP_asymmetries_bm_point}
\end{figure}

In Fig.~\ref{fig:observ_dep_bm_point}, we show $P_2, P_3, P_4', P_5'$ and $P_6'$ as a function of the NP Wilson coefficients. In Fig.~\ref{fig:observ_dep_CP_asymmetries_bm_point}, we show the same for $A_{3,4,5,6s,8,9}$. The gray bars indicate the experimental values of Tables~\ref{tab:exp_constraints_cpavg} and~\ref{tab:exp_constraints_cpasymm}.

\begin{figure}[h]
    \centering
    \subfloat[]{\includegraphics[width=0.3\linewidth]{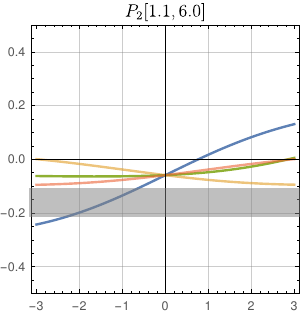}}
    \hfill
    \subfloat[]{\includegraphics[width=0.3\linewidth]{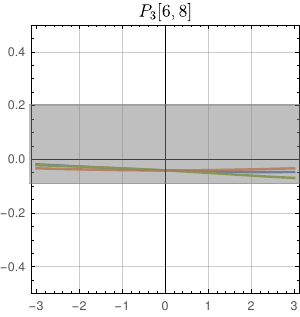}}
    \hfill
    \subfloat[]{\includegraphics[width=0.3\linewidth]{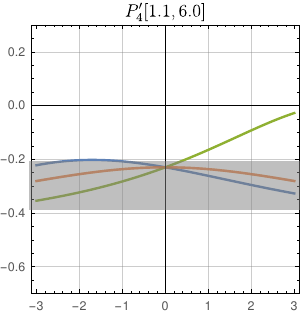}}
    \hfill\\
    \subfloat[]{\includegraphics[width=0.3\linewidth]{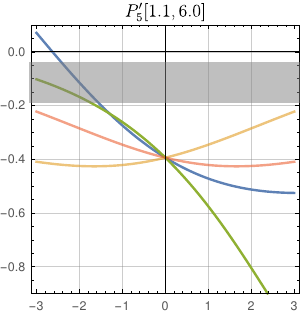}}
    \hfill
    \subfloat[]{\includegraphics[width=0.3\linewidth]{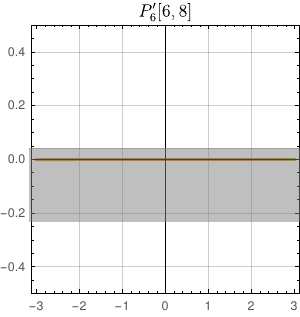}}
    \hfill
    \caption{Sensitivity of CP-averaged observables on the variation of the complex Wilson coefficients. Here we set $\cplus_9$ and $\cplus_{10}$ to the values in \eqref{eq:c9plus_c10plus_bm} and then turn on the real and imaginary parts of $\mathcal{C}_9^{- \rm NP}$ and $\mathcal{C}_{10}^{- \rm NP}$ independently. The grey bars indicate the experimental measurements in Table~\ref{tab:exp_constraints_cpavg}.}
    \label{fig:observ_dep_bm_point}
\end{figure}

\newpage
\section{Time-dependent CP asymmetries}
\label{sec:timedep}

\begin{figure}[th]
    \centering
    \subfloat[]{\includegraphics[width=0.3\linewidth]{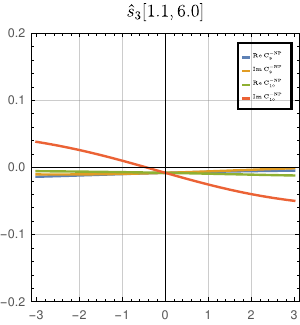}}
    \hfill
    \subfloat[]{\includegraphics[width=0.3\linewidth]{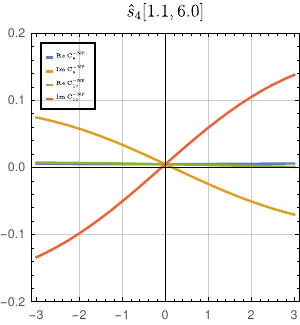}}
    \hfill
    \hfill
    \subfloat[]{\includegraphics[width=0.3\linewidth]{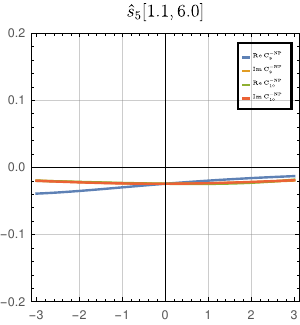}}\\
    \hfill
    \subfloat[]{\includegraphics[width=0.3\linewidth]{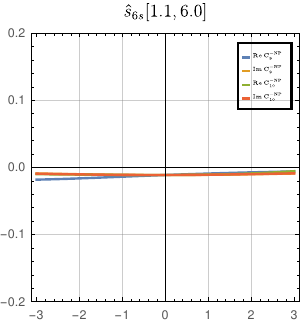}}
    \hfill
    \subfloat[]{\includegraphics[width=0.3\linewidth]{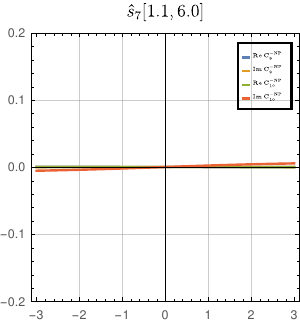}}
    \hfill
    \subfloat[]{\includegraphics[width=0.3\linewidth]{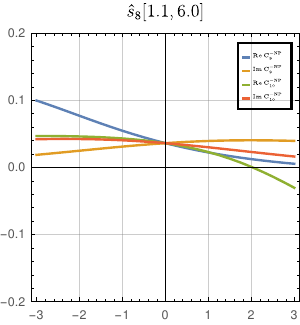}}
    \hfill
    \subfloat[]{\includegraphics[width=0.3\linewidth]{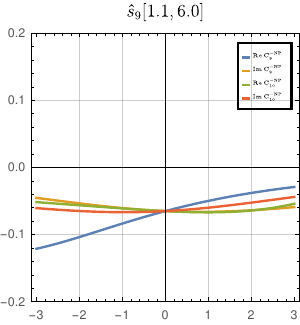}}
    \hfill
    \caption{Dependences of the $s_i$ observables, accessible through a time-dependent analysis, on the real and imaginary parts of $\cminus_9$ and $\cminus_{10}$. Here we consider all the observables in the $[1.1,6.0]$ GeV$^2$ bin.}
    \label{fig:time_dep_obs_deps}
\end{figure}
The definitions of the angular coefficients $s_i$ and $h_i$ in terms of transversity amplitudes can be found in \cite{Descotes-Genon:2015hea}. We define the following normalized angular coefficients, similar to those used for $B \to K\ell^+\ell^-$ in \cite{Descotes-Genon:2020tnz}:
\begin{equation}
    \hat s_i \equiv \frac{s_i}{\Gamma + \bar \Gamma} \ , \quad \hat h_i \equiv \frac{h_i}{\Gamma + \bar \Gamma} \ .
    \end{equation}

\begin{figure}[h!]
    \centering
    \subfloat[]{\includegraphics[width=0.3\linewidth]{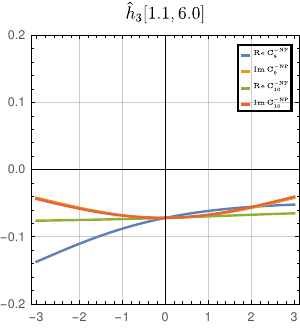}}
    \hfill
    \hfill
    \subfloat[]{\includegraphics[width=0.3\linewidth]{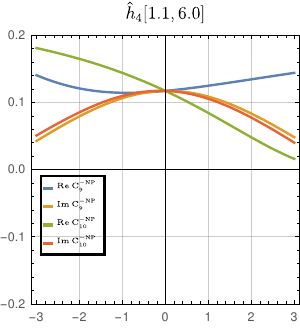}}
    \hfill
    \subfloat[]{\includegraphics[width=0.3\linewidth]{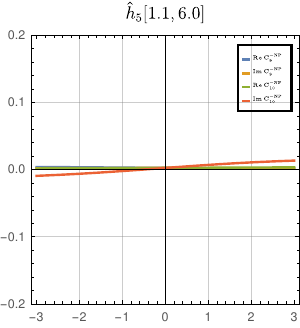}}\\
    \subfloat[]{\includegraphics[width=0.3\linewidth]{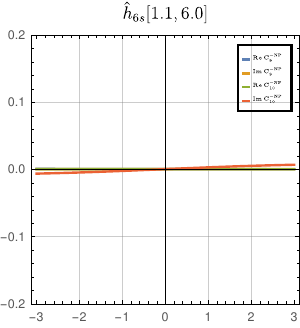}}
    \hfill
    \hfill
    \subfloat[]{\includegraphics[width=0.3\linewidth]{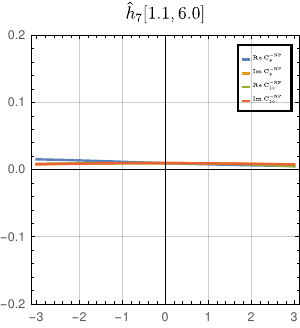}}
    \hfill
    \hfill
    \subfloat[]{\includegraphics[width=0.3\linewidth]{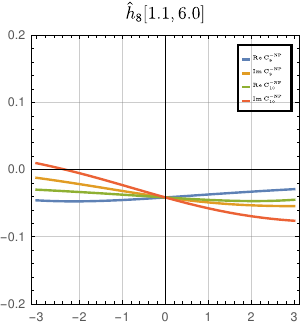}}\\
    \subfloat[]{\includegraphics[width=0.3\linewidth]{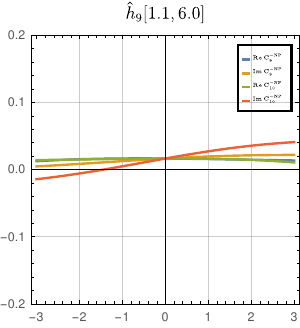}}
    \hfill
    \caption{Dependences of the $h_i$ observables, accessible through a time-dependent analysis, on the real and imaginary parts of $\cminus_9$ and $\cminus_{10}$. Here we consider all the observables in the $[1.1,6.0]$ GeV$^2$ bin.}
    \label{fig:time_dep_obs_deph}
\end{figure}

For completeness, we show the dependences of $\hat{s}_i$ and $\hat{h}_i$ on in Fig.~\ref{fig:time_dep_obs_deps} and Fig.~\ref{fig:time_dep_obs_deph}, respectively.  As before, we set $\cplus_9$ and $\cplus_{10}$ to the values in \eqref{eq:c9plus_c10plus_bm} and then turn on the real and imaginary parts of $\mathcal{C}_9^{- \rm NP}$ and $\mathcal{C}_{10}^{- \rm NP}$ independently. 

\newpage
\bibliography{refs.bib}
\end{document}